\documentclass[preprint]{elsarticle}

\usepackage{filecontents}
\usepackage{epstopdf}
\usepackage{caption}
\usepackage{subcaption}
\usepackage{tikz}
\usepackage{overpic}
\usepackage{import}
\usepackage{here} 
\usepackage{pgf}
\usepackage[utf8]{inputenc}

\usepackage{booktabs}
\usepackage{enumitem}
\usepackage{geometry}
\usepackage{lineno}

\usepackage[version=3]{mhchem}
\usepackage{mathrsfs}

\usepackage{bibspacing}

\usepackage{multicol}
\usepackage{nomencl}
\usepackage{etoolbox}

\usepackage{hyperref}

\usepackage{graphicx}
\usepackage{booktabs}
\usepackage{multirow}
\usepackage{microtype}
\usepackage{longtable}
\usepackage{tabularx}
\usepackage{doi}
\usepackage{dblfloatfix}
\usepackage{soul}
\usepackage{siunitx}
\usepackage{framed}
\usepackage{nomencl}
\usepackage{derivative}
\usepackage[export]{adjustbox}
\usepackage{float}
\usepackage[T1]{fontenc}
\usepackage{xcolor}
\usepackage{tcolorbox}\tcbuselibrary{breakable}
\usepackage{amsmath}
\usepackage{amssymb}

\journal{Combustion and Flame}

\begin{document}

\begin{frontmatter}

\title{
Assessment of the Flamelet Generated Manifold method with preferential diffusion modelling for the prediction of partially premixed hydrogen flames}

\author{E. J. Pérez-Sánchez\corref{cor1}}
\cortext[cor1]{Corresponding author}
\ead{eduardo.perez@bsc.es}
\author{E. M. Fortes}
\author{D. Mira}
\address{Barcelona Supercomputing Center (BSC), Plaza Eusebi Güell 1-3, 08034, Barcelona, Spain}

\begin{abstract}

This study presents a systematic analysis of the capabilities of a flamelet model based on Flamelet Generated Manifolds (FGM) to reproduce preferential diffusion effects in partially premixed hydrogen flames. Detailed transport effects are accounted for by including a mixture-averaged transport model
when building the flamelet database. This approach adds new terms to the transport equations of the controlling variables in the form of diffusive fluxes where the coefficients can be computed from the information contained in the manifold and saved in the flamelet database. The manifold is constructed from the solution of a set of premixed unstretched adiabatic one-dimensional flames with mixture-averaged transport for a range of mixture fraction within the flammability range. Special attention is given to the numerical aspects related to the construction of the chemical manifold to reduce the numerical error in evaluating the new terms derived from the preferential diffusion of certain species. Finally, a systematic application of the method to simulate laminar hydrogen flames in various canonical configurations is presented from premixed to stratified flames, including the case of a triple flames with different mixing lengths. The results demonstrate that the method describes accurately the flame structure and propagation velocities at a low cost, showing a remarkable agreement with the detailed chemistry solutions for flame structure and propagation velocity.
\end{abstract}


\end{frontmatter}

\section{Introduction}

The need of designing clean and green combustion systems with net zero CO$_2$-emissions has impelled the research in hydrogen combustion due to its high mass energy content, absence of carbon-derived pollutants emissions like soot and potentially low NOx formation when operated in lean conditions.
However, the high flame speed together with its relative low density energy along with effect of preferential diffusion can lead to a variety of phenomena, including thermoacoustics and
thermodiffusive instabilities, that can have a direct impact on the engine design and operation.

Preferential diffusion introduces a set of topological changes in the flame that affect both its morphology and dynamics, and whose understanding has been a target for the scientific community for decades. Preferential diffusion effects induce a local variability in the mixture fraction which, for low enough equivalence ratios, leads to non-conventional flame configurations such as the flame tip opening of lean hydrogen-air bunsen flames \cite{Mizobuchi2019}. Similarly, preferential diffusion impacts on the morphology of the flame by creating cellular structures like polyhedral flames in hydrogen lean premixed flames as shown in \cite{Shi2023} and cellular structures \cite{Matalon2009}. Analytical expressions were obtained for the Markstein length and dispersion relationship when considering preferential diffusion \cite{Matalon1983} and \cite{Matalon2003}. It was found that a destabilizing behaviour appears for the lean hydrogen mixtures (thermo-diffusive instability).

Also detailed numerical simulations 
have been conducted to understand the influence of preferential diffusion in hydrogen flames~\cite{Berger2022_partI,Berger2022_partII,Vance2022,Vance2023}. These studies have shown the
connections between the vortex-flame interactions and the flame surface and global burning velocity \cite{Kadowaki2005}, evidencing the influence of the equivalence ratio, pressure and temperature on the linear and non-linear regimes for hydrogen flames \cite{Berger2022_partI,Berger2022_partII}.

Regarding the modelling of preferential diffusion, different modellings strategies have been presented in the literature as explained in the following. The main difficulty underlays in capturing the diffusive flux involving species with different diffusivities across the flame front, while using reliable methods at reduced computational cost.

Among the different approaches in the literature to describe turbulent combustion~\cite{MIRA2022}, the Flamelet Generated Manifold (FGM) method~\cite{vanOijen2000,VANOIJEN201630} has demonstrated an excellent trade-off between accuracy and computational cost. It has been widely applied to a range of different applications from gaseous~\cite{Govert2018,Both_CAF_2020,mira2020numerical,Pachano2023} to spray flames~\cite{BENAJES2022111730,Mira2021_icef}. The FGM is based on the flamelet concept and assumes that the thermochemical states of the flame mainly remain on a manifold, typically constructed from a set of one-dimensional laminar flames. The manifold only depends on a reduced set of independent variables, commonly known as controlling variables, which are denoted as $\mathbf{\mathcal{Y}}=(\mathcal{Y}_1,\ldots,\mathcal{Y}_{N_c})$. These control variables $\mathbf{\mathcal{Y}}$ are usually defined by linear combination of species mass fractions \cite{Govert2015}. For the case that species diffusion is governed by Fick's law  with unity Lewis number, diffusive transport terms can be easily derived that only depend on the specific controlling variable.

Nevertheless, when introducing differential diffusion effects, the evaluation of the terms related to the diffusive transport becomes more complex. Different strategies have been proposed to extend the capabilities of tabulated chemistry methods to model preferential diffusion. First works including detailed transport in tabulated chemistry methods considered the calculation and storage of effective Lewis numbers for each controlling variable in the manifold space \cite{Vreman2009}. Those were obtained from the Lewis numbers of the individual species and the spatial derivatives of the mass fractions in the one-dimensional flames. It was emphasized, however, the non-negligible dependence of the results on the choice of the controlling variables. In later works, preferential diffusion was accounted for by introducing new terms in the control variable transport equations when evaluating the diffusive fluxes \cite{deSwart2010,Dinesh2013,Donini2015}. Based on the dependencies imposed by the manifold and after applying the chain rule one obtains: 

\begin{equation}
\sum_{i=1}^{N_s} \alpha_{ij} D_i \nabla Y_i = \sum_{k=1}^{N_c} \left( \sum_{i=1}^{N_s}\alpha_{ij} D_i \frac{\partial Y_i}{\partial \mathcal{Y}_k}\right) \nabla \mathcal{Y}_k = \sum_{k=1}^{N_c} \Gamma_{jk} \nabla \mathcal{Y}_k,
\end{equation}

where $\mathcal{Y}_k$ denotes the set of $N_c$ controlling variables, $N_s$ is the number of species and $\alpha_{ij}$ is the coefficient of the i-th species in the definition of the j-th controlling variable. As the definition of $\Gamma_{jk}$ only depends on the definition of the controlling variables and the internal flame structure, their values can be tabulated as a function of the controlling variables. This derivation only affects the diffusion terms and introduces crossed contributions between control variables (terms related to $j \neq k$).

This approach has been extensively used in a series of studies where premixed flames were simulated using hydrogen/methane mixtures \cite{deSwart2010}, pure methane with stratification and heat losses \cite{Donini2015} and pure hydrogen and $\mathrm{H_2/CO}$ syngas flames \cite{Dinesh2013}. However, in all of them only the contribution from the progress variable ($\Gamma_{1i} \nabla \mathcal{Y}_1$) where subscript 1 refers to the progress variable was retained. Despite this approximation, the main characteristics of the flame were successfully reproduced. More recently this method was applied to Large Eddy Simulations \cite{Donini2017,Almutairi2023}. A full formulation considering cross terms was applied in \cite{Abtahizadeh2015,Nicolai2022} with constant non-unity Lewis numbers to model hydrogen enriched methane-based fuel mixtures.

An alternative approach based on the Flamelet Progress Variable (FPV) model was devised in \cite{Bottler2022} and successfully applied to forced ignition and flame propagation in hydrogen-air mixtures. A set of species mass fractions were transported to reconstruct the controlling variables in order to access the manifold. The method was applied to the simulation of cylindrical and spherical expanding flames as well as a counterflow flame showing accuracy in the prediction of flame characteristics. In contrast, in \cite{Wen2022_part_I,Wen2022_part_II} investigations of ultra-lean premixed hydrogen flame with significant thermodiffusive instabilities at atmospheric and high pressures were performed by considering an FPV model including curvature and strain. Finally, both models were compared in \cite{Bottler2023} for the simulation of expanding spherical flames showing that the last formulation was more convenient to describe the instabilities growth during the linear regime.

Apart from these investigations, a new variant of tabulated chemistry with detailed transport was originally developed by Regele et al.~\cite{Regele2013} in the frame of the FPV model. It includes a new correction term for the mixture fraction equation derived from the fluxes of the oxygen and the fuel. While in the initial derivation a strong restriction based on irreversible one-step chemistry was introduced, this hypothesis was then relaxed to broaden the applicability of the model to complex chemical schemes. In this case, only the Lewis number for the fuel was considered different to unity. The approach was shown to correctly predict the dynamics of symmetrical inward and outward propagating spherical hydrogen-air and propane flames. The model was later improved in \cite{Schlup2019} where a fully mixture-averaged model was considered and thermal diffusion effects were also included to capture local and global properties of curved flames.

Finally, an alternative approach to include preferential diffusion effects in FGM was presented in the recent work by Mukundakumar et al.~\cite{Mukundakumar2021}. The model is based on commuting the order of the operations for the diffusive transport terms of the controlling variables after imposing constant Lewis numbers and grouping the contributions into new coefficients that are tabulated in the manifold. The model is shown to improve the predictions of the previous model~\cite{Donini2015}, but it is limited to constant Lewis numbers, as its extension to mixture-averaged transport is not straightforward.

However, despite the substantial progress and advancements carried out towards incorporating preferential diffusion effects in tabulated chemistry methods, it is still unclear the limitations of such approaches to accurately reproduce the flame structure and dynamics in the case of partially premixed flames. Starting from the the work by~\cite{deSwart2010,Donini2015}, this paper presents a systematic analysis and evaluation of the capabilities of a flamelet model based on the tabulation of one-dimensional unstretched laminar premixed flames with a mixture-averaged transport considering a full formulation. Moreover, focus is given to the calculation of the resulting coefficients and the interpolation algorithm in order to reduce the numerical errors in the manifold tabulation. Then, the influence of the degree of stratification is widely examined in canonical flame configurations involving different levels of premixing.

The paper is structured as follows: first, a description of the model is given. Attention is devoted to not only the formulation of the model but also to the numerical aspects related to the construction of the manifold. The model is also validated by reproducing one-dimensional flames. Second, results for a stratified flame with 
a limited range of mixture fraction and a triple flame configuration are thoroughly analysed. Discussion are presented in terms of the capabilities of the FGM model to reproduce the reference results from detailed chemistry. Conclusions and future work close the work.

\section{Model description}\label{sec:Model-description}

In this section, a detailed description of the main assumptions and derivation of the flamelet model with mixture-averaged transport is presented with special emphasis on the aspects related to the tabulation strategy. While different flamelet methods with preferential diffusion effects exist in the literature~\cite{deSwart2010,Regele2013,Mukundakumar2021,Wen2022_part_I,Wen2022_part_II,Nicolai2022}, the current model is based on the concepts presented in \cite{deSwart2010,Donini2015} but using a mixture-averaged diffusion model, which is a more general and accurate approach than those based on constant Lewis numbers.

\subsection{Theoretical framework}

The mixture-averaged diffusion model approximates the diffusion fluxes for species as:

\begin{equation}\label{diff_eq_MA}
\mathbf{V_k}X_k = - D_k \nabla X_k \quad \quad k=1,\ldots,N_s,
\end{equation}

being $\mathbf{V_k}$ the diffusion velocity for $k$-th species, $X_k$ its molar fraction and $N_s$ the number of species (vectors are denoted with bold type). The diffusion coefficient $D_k$ can be determined as:

\begin{equation}\label{Dk_MA}
D_k = \frac{1-Y_k}{\sum_{\substack{j=1 \\ j \neq k}}^{N_s} X_j / \mathcal{D}_{jk}}
\end{equation}

with $Y_k$ being the mass fraction for the $k$-th species and $\mathcal{D}_{jk}$ the binary diffusion coefficient of the $j$-th species into the $k$-th species. As transport equations are usually solved for species mass fractions, it is more convenient to rewrite Eq. (\ref{diff_eq_MA}) in terms of $Y_k$, considering that $Y_k = X_k W_k / W$ with $W_k$ and $W$ the molecular weights for the $k$-th species and the mixture, respectively. The  diffusion flux is then expressed as:

\begin{equation}
\mathbf{V_k}Y_k = - D_k \frac{W_k}{W} \nabla X_k=-D_k\nabla Y_k -D_kY_k\frac{\nabla W}{W}.
\end{equation}

As the mixture averaged model does not guarantee that the net diffusion flux is balanced out for all the species, a velocity correction is usually introduced to ensure mass conservation, leading to:

\begin{equation}
\sum_{j=1}^{N_s} (\mathbf{V_k}+\mathbf{V_c})Y_k=0.
\end{equation}

The diffusion flux for the $k$-th species $\mathbf{j}_k = (\mathbf{V}_k+\mathbf{V_c})Y_k$ finally takes the form:

\begin{equation}\label{diff_eq_MA_total}
\begin{split}
\mathbf{j}_k = & -D_k \nabla Y_k
-D_k Y_k\frac{\nabla W}{W} +Y_k\sum_{j=1}^{N_s}D_j\nabla Y_j + \frac{\nabla W}{W} Y_k\sum_{j=1}^{N_s}D_j Y_j.
\end{split}
\end{equation}

Despite it is well-known that the contribution due to the correction velocity and the molecular weight can be small~\cite{Regele2013}, these terms are retained in this formulation to ensure mass conservation and consistency between the flamelet solution and the governing equations in the FGM method. Besides, the cost of their evaluation is negligible and can be introduced in the flamelet manifold without any additional complexity.

The proposed flamelet model is based on the FGM concept, which assumes that the chemical time scales are faster than
the flow scales, and the flame structure can be defined by a manifold in a low order space using appropriate control variables. According to the FGM, such manifold is well-represented by the thermochemical structure of a set of one-dimensional flames (flamelet concept) \cite{vanOijen2000,VANOIJEN201630} and this approach is used here to include mixture fraction variations.

As a consequence, any dependent variable (mass fractions, mean molecular weight, etc.) under the FGM assumptions is a function of the controlling variables given by the $N_c$-tuple $(\mathcal{Y}_1, \ldots, \mathcal{Y}_{N_c})$, existing an injective application from the thermochemical states into the space of controlling variables. Hence, expressing the species mass fractions as function of the control variables to Eq. (\ref{diff_eq_MA_total}), the diffusive flux can be written as:

\begin{equation}\label{chain_rule_flux}\begin{split}
\mathbf{j}_k = 
- \sum_{i=1}^{N_c} \Biggl( & D_k \frac{\partial Y_k}{\partial \mathcal{Y}_i} +
\frac{D_kY_k}{W} \frac{\partial W}{\partial \mathcal{Y}_i} 
- Y_k\sum_{j=1}^{N_s}D_j \frac{\partial Y_j}{\partial \mathcal{Y}_i} - 
\\& -\frac{Y_k}{W} \frac{\partial W}{\partial \mathcal{Y}_i} \sum_{j=1}^{N_s}D_jY_j \Biggr) \nabla \mathcal{Y}_i = 
- \sum_{i=1}^{N_c} \Lambda_{Y_k,\mathcal{Y}_i} \nabla \mathcal{Y}_i.
\end{split}
\end{equation}

The terms in brackets $\Lambda_{Y_k,\mathcal{Y}_i}$ represent the contribution of each of the controlling variable into the diffusive flux for the $k$-th species. These quantities are expressed in composition space, so there is no dependency from the computational domain and their spatial dependence is entirely contained in the gradients of the controlling variables $\nabla \mathcal{Y}_i$. Hence, these quantities can be obtained in a pre-processing stage and then incorporated into the flamelet manifold. For many applications the manifold space can be determined by three control variables $(\mathcal{Y}_1,\mathcal{Y}_2,\mathcal{Y}_3)=(Y_c,Z,h)$ (progress variable, mixture fraction and enthalpy, respectively). Derivations for those quantities are presented in the following.

Considering that the transport equation for the $k$-th species takes the form:

\begin{equation}\label{eq:transport_Yk}
    \rho \frac{\partial Y_k}{\partial t} + \rho \boldsymbol{u} \cdot \nabla Y_k + \nabla \cdot (\rho \,\mathbf{j}_k)  = \rho \, \dot{\omega}_k,
\end{equation}

where $\rho$, $\boldsymbol{u}$ and $\dot{\omega}_k$ are the density, velocity and reaction source term for the $k$-th species, respectively, a transport equation for $Y_c$ can be defined.

The progress variable $Y_c$ describes the degree of advancement of the combustion process and is usually considered as a linear combination of species mass fractions $Y_c=\sum_{k=1}^{N_s} \alpha_k Y_k$. Adding the transport equations for the species weighed by the coefficients $\alpha_k$, a transport equation for $Y_c$ can be obtained:

\begin{equation}\label{eq_Yc}
\rho \frac{\partial Y_c}{\partial t} + \rho \boldsymbol{u} \cdot \nabla Y_c + \nabla \cdot \left(\rho \sum_{k=1}^{N_s} \alpha_k \, \mathbf{j}_k \right) = \rho \sum_{k=1}^{N_s} \alpha_k \, \dot{\omega}_k.
\end{equation}

The diffusion flux contribution for $Y_c$ can be easily composed from Eq.~(\ref{chain_rule_flux}) and rearranged as:

\begin{equation}
 \sum_{k=1}^{N_s} \alpha_k \, \mathbf{j}_k = -  \sum_{i=1}^{N_c} \Gamma_{Y_c,\mathcal{Y}_i} \nabla \mathcal{Y}_i,
\end{equation}
where $\Gamma_{Y_c,\mathcal{Y}_i}$ denotes the contribution of the controlling variable $\mathcal{Y}_i$ on $Y_c$:

\begin{equation}
\Gamma_{Y_c,\mathcal{Y}_i} = \sum_{k=1}^{N_s} \alpha_k \Lambda_{Y_k,\mathcal{Y}_i}.
\end{equation}

Finally, one can write the transport equation for the progress variable including the diffusive fluxes of all the species for each control variable as:

\begin{equation}\label{eq_Yc_final}
\rho \frac{\partial Y_c}{\partial t} + \rho \boldsymbol{u} \cdot \nabla Y_c  = \sum_{i=1}^{N_c} \nabla \cdot (\rho \, \Gamma_{Y_c,\mathcal{Y}_i} \, \nabla \mathcal{Y}_i ) + \rho \, \dot{\omega}_{Y_c},
\end{equation}
with $\dot{\omega}_{Y_c} = \sum_{k=1}^{N_s} \alpha_k \, \dot{\omega}_k$. Note that the fluxes $\nabla \cdot (\rho \, \Gamma_{Y_c,\mathcal{Y}_i} \, \nabla \mathcal{Y}_i )$ do not necessarily correspond to a diffusion term, since coefficients $\Gamma_{Y_c,\mathcal{Y}_i}$ can be negative.

In a similar way, the transport equation for the total enthalpy $h$ (sensible plus chemical) can be written as:

\begin{equation}\label{eq_h}
\rho \frac{\partial h}{\partial t} + \rho \boldsymbol{u} \cdot \nabla h + \nabla \cdot (\rho \, \mathbf{j_h} )= 0,
\end{equation}

where the dissipation due to viscous forces ($\Phi_{ij}= \tau_{ij} \partial u_i / \partial x_j$), pressure  term, and body forces contribution are neglected. After applying similar considerations than in Eq. (\ref{chain_rule_flux}), the diffusive flux of enthalpy $\mathbf{j_h}$ is given by:

\begin{equation}\label{eq_jh}
\begin{split}
 \mathbf{j_h} &=-\frac{\lambda}{\rho} \nabla T +  \sum_{k=1}^{N_s} (\mathbf{V}_k+\mathbf{V_c})Y_k h_k
=-\sum_{i=1}^{N_c} \left(\frac{\lambda}{\rho} \frac{\partial T}{\partial \mathcal{Y}_i}+
 \sum_{k=1}^{N_s}\Lambda_{Y_k,\mathcal{Y}_i} h_k \right) \nabla \mathcal{Y}_i
\\&=- \sum_{i=1}^{N_c} \Gamma_{h,\mathcal{Y}_i} \nabla \mathcal{Y}_i,
\end{split}
\end{equation}
where $\lambda$ denotes the thermal conductivity and $h_k$ is the specific enthalpy for the k-th species. The coefficient $\Gamma_{h,\mathcal{Y}_i}$ is given by:

\begin{equation}
\begin{split}
\Gamma_{h,\mathcal{Y}_i} 
= \frac{\lambda}{\rho} \frac{\partial T}{\partial \mathcal{Y}_i}
+ \sum_{k=1}^{N_s}\Lambda_{Y_k,\mathcal{Y}_i} h_k.
\end{split}
\end{equation}

Finally, the enthalpy equation can be rewritten as:

\begin{equation}\label{eq_h_final}
\rho \frac{\partial h}{\partial t} + \rho \boldsymbol{u} \cdot \nabla h = \sum_{i=1}^{N_c} \nabla \cdot (\rho \, \Gamma_{h,\mathcal{Y}_i} \nabla \mathcal{Y}_i ).
\end{equation}
 
Finally, the equation for mixture fraction is considered. Bilger's definition for the mixture fraction $Z_p$ associated to an element $p$ is:

\begin{equation}\label{definition_mixt_fract_elem}
Z_p=W_p\sum_{k=1}^{N_s} a_{kp}\frac{Y_k}{W_k},
\end{equation}

with $W_p$ the molecular weight for the element $p$ and $a_{kp}$ the number of atoms of element $p$ in the $k$-th species. For a generic hydrocarbon of composition $\mathrm{C_mH_n}$ and using these definitions, the mixture fraction $Z$ can be obtained as:

\begin{equation}\label{definition_mixt_fract}
Z = \frac{\frac{Z_C}{mW_C} \,+ \, \frac{Z_H}{nW_H} \, + \, 2\frac{Y_{O_2,2}-Z_O}{\nu_{\mathrm{O_2}}W_{\mathrm{O_2}}}}
{\frac{Z_{C,1}}{mW_C} \,+ \, \frac{Z_{H,1}}{nW_H} \, + \, 2\frac{Y_{O_2,2}}{\nu_{\mathrm{O_2}}W_{\mathrm{O_2}}}}
= K_Z \left(\frac{Z_C}{mW_C} \,+ \, \frac{Z_H}{nW_H} \, + \, 2\frac{Y_{O_2,2}-Z_O}{\nu_{\mathrm{O_2}}W_{\mathrm{O_2}}} \right)
\end{equation}

with $Z_C$, $Z_H$ and $Z_O$ being the mixture fractions for atomic carbon, hydrogen and oxygen, respectively, $\nu_{\mathrm{O_2}}$ the stoichiometric coefficient for oxygen and subscripts 1 and 2 denote the fuel and oxidizer streams, respectively. Finally, $Y_{O_2,2}$ denotes the mass fraction for molecular oxygen in the oxidant reactant. The inverse of the denominator in Eq. (\ref{definition_mixt_fract}), which only depends on the composition in the fuel and oxidant streams, is compacted in the constant $K_Z$.

Linearly combining the species transport equations according to the definition from Eq. (\ref{definition_mixt_fract_elem}), the transport equation for $Z_p$ reads:

\begin{equation}\label{eq_Zp}
\rho \frac{\partial Z_p}{\partial t} + \rho \boldsymbol{u} \cdot \nabla Z_p + \nabla \cdot \left(\rho W_p\sum_{k=1}^{N_s}\frac{a_{kp}}{W_k} \mathbf{j}_k \right) = 0,
\end{equation}

where the diffusive flux can be written using the same notation as the progress variable and enthalpy given above as:

\begin{equation}
\begin{split}
\mathbf{j_{Z_p}} &= W_p\sum_{k=1}^{N_s}\frac{a_{kp}}{W_k} \mathbf{j}_k 
=-  \sum_{i=1}^{N_c} \left[ \sum_{k=1}^{N_s}\frac{a_{kp}}{W_k}\Lambda_{Y_k,\mathcal{Y}_i} \right] W_p \nabla \mathcal{Y}_i
= - \sum_{i=1}^{N_c} \Gamma_{Z_p,\mathcal{Y}_i}\nabla \mathcal{Y}_i.
\end{split}
\end{equation}

In turn, the transport equation for mixture fraction $Z$ can be easily deduced by linear combination of the transport equations for $Z_p$ (Eq. (\ref{eq_Zp})) according to the definition of Eq. (\ref{definition_mixt_fract}), yielding:

\begin{equation}\label{eq_Z}
\rho \frac{\partial Z}{\partial t} + \rho \boldsymbol{u} \cdot \nabla Z + \nabla \cdot \left(\rho \, \mathbf{j_Z} \right) = 0,
\end{equation}

where the diffusive flux for mixture fraction $\mathbf{j_Z}$ is given by:

\begin{equation}\label{eq_jZ}
\mathbf{j_Z} =  K_Z \left( \frac{\mathbf{j_{Z_C}}}{mW_C} \,+ 
\, \frac{\mathbf{j_{Z_H}}}{nW_H} \, - \, 2\frac{\mathbf{j_{Z_O}}}{\nu_{\mathrm{O_2}}W_{\mathrm{O_2}}} \right)
= - \sum_{i=1}^{N_c} \Gamma_{Z,\mathcal{Y}_i} \nabla \mathcal{Y}_i,
\end{equation}

being $\Gamma_{Z,\mathcal{Y}_i}$ given by:

\begin{equation}
\Gamma_{Z,\mathcal{Y}_i} 
= K_Z \left(\frac{\Gamma_{Z_C,\mathcal{Y}_i}}{mW_C} \,+ 
\, \frac{\Gamma_{Z_H,\mathcal{Y}_i}}{nW_H} \, - \, 2\frac{\Gamma_{Z_O,\mathcal{Y}_i}}{\nu_{\mathrm{O_2}}W_{\mathrm{O_2}}}\right).
\end{equation}

Therefore, finally it can be written:

\begin{equation}\label{eq_Z_final}
\rho \frac{\partial Z}{\partial t} + \rho \boldsymbol{u} \cdot \nabla Z = \sum_{i=1}^{N_c} \nabla \cdot (\rho \, \Gamma_{Z,\mathcal{Y}_i} \, \nabla \mathcal{Y}_i ).
\end{equation}

The process can be generalized for any control variable $\mathcal{Y}_i$ taking the equation, finally, the form:

\begin{equation}\label{eq:eq_Y_control_general}
\rho \frac{\partial \mathcal{Y}_i}{\partial t} + \rho \boldsymbol{u} \cdot \nabla \mathcal{Y}_i = \sum_{j=1}^{N_c} \nabla \cdot (\rho \, \Gamma_{\mathcal{Y}_i,\mathcal{Y}_j} \, \nabla \mathcal{Y}_j ) + S_{\mathcal{Y}_i},
\end{equation}

where $S_{\mathcal{Y}_i}$ represents the chemical source term for $\mathcal{Y}_i$ (if exists). 

To formally separate the different effects, the unity Lewis contribution is separated from the preferential diffusion effects by defining the following coefficients:

\begin{equation}\label{eq:coeff_stab}
\Gamma^{'}_{\mathcal{Y}_i,\mathcal{Y}_j} =
\begin{cases}
\Gamma_{\mathcal{Y}_i,\mathcal{Y}_i} - D_{th} & i = j \\
\Gamma_{\mathcal{Y}_i,\mathcal{Y}_j} & i \neq j
\end{cases}
\end{equation}

where the thermal diffusivity $D_{th}$ is obtained by $D_{th}=\lambda / (\rho c_p)$, being $c_p$ the specific heat at constant pressure of the mixture. Finally, the transport equation for $\mathcal{Y}_i$ is written as:

\begin{equation}\label{eq_Yi_general_stab}
\rho \frac{\partial \mathcal{Y}_i}{\partial t} + \rho \boldsymbol{u} \cdot \nabla \mathcal{Y}_i = \nabla \cdot (\rho \, D_{th} \, \nabla \mathcal{Y}_i) +
\sum_{j=1}^{N_c} \nabla \cdot (\rho \, \Gamma^{'}_{\mathcal{Y}_i,\mathcal{Y}_j} \, \nabla \mathcal{Y}_j ) + S_{\mathcal{Y}_i}.
\end{equation}

In all the cases, the transport equation contains a set of coefficients $\Gamma^{'}_{\mathcal{Y}_i,\mathcal{Y}_j}$ (dependent on the $N_c$-tuple defined by the controlling variables), which expresses the contribution of each control variable $\mathcal{Y}_j$ into the transport equation for $\mathcal{Y}_i$ due to preferential diffusion effects. Note that the evolution of one controlling variable is influenced by the others (cross terms $\nabla \cdot (\rho \, \Gamma^{'}_{\mathcal{Y}_i,\mathcal{Y}_j} \, \nabla \mathcal{Y}_j ) \quad i \neq j$) due to preferential diffusion effects. However, depending on the flame conditions, some of these cross terms can be neglected and simpler models can be easily derived \cite{deSwart2010,Dinesh2013,Donini2015}. Furthermore, notice that the proposed modelling framework guarantees the conservation of the controlling variables due to the addition of the preferential diffusion terms as diffusive fluxes $\sum_{j=1}^{N_c} \nabla \cdot (\rho \, \Gamma^{'}_{\mathcal{Y}_i,\mathcal{Y}_j} \, \nabla \mathcal{Y}_j)$. This can be easily demonstrated by integrating Eq. (\ref{eq:eq_Y_control_general}) and using the divergence theorem.

In addition, this representation of the diffusive flux enables a simple coupling with the flamelet method, since the coefficients $\Gamma^{'}_{\mathcal{Y}_i,\mathcal{Y}_j}$ can be directly computed from the flamelet solutions and then, be included in the chemical manifold. However, these terms contain derivatives in the phase space (defined by the set $(\mathcal{Y}_1, \ldots, \mathcal{Y}_{N_c})$) but, due to preferential diffusion, the flamelet solutions describe curved trajectories in such space (in general, all the controlling variables will change along one given flame), so their calculation can bring some difficulties. The next subsection describes a simple strategy that enables an efficient and accurate tabulation of the $\Gamma^{'}_{\mathcal{Y}_i,\mathcal{Y}_j}$ fields (in phase space), so the interpolation errors during the tabulation process are minimized.

\subsection{Construction of the flamelet manifold with preferential diffusion}

In this section, some considerations related to the construction of the chemical manifold are given. In this study, the manifolds are generated from a set of one-dimensional unstretched adiabatic laminar premixed flames at different equivalence ratios covering the flammability range. Therefore, two control variables are used to describe the manifold space, namely, the progress variable $Y_c$ and the mixture fraction $Z$ ($\psi=\psi(Y_c,Z)$).

Rewriting Eqs. (\ref{eq_Yc_final}) and (\ref{eq_Z_final}) and using the coefficients defined in Eq. (\ref{eq:coeff_stab}), the transport equations for $Y_c$ and $Z$ are given by:

\begin{equation}\label{eq_solve_Yc}
\rho \frac{\partial Y_c}{\partial t} + \rho \boldsymbol{u} \cdot \nabla Y_c = \nabla \cdot (\rho \, D_{th} \, \nabla Y_c +\rho \, \Gamma^{'}_{Y_c,Y_c} \, \nabla Y_c  + \rho \, \Gamma^{'}_{Y_c,Z} \, \nabla Z ) +
\rho \, \dot{\omega}_{Y_c},
\end{equation}

\begin{equation}\label{eq_solve_Z}
\rho \frac{\partial Z}{\partial t} + \rho \boldsymbol{u} \cdot \nabla Z = \nabla \cdot (\rho \, D_{th} \, \nabla Z +\rho \, \Gamma^{'}_{Z,Y_c} \, \nabla Y_c  + \rho \, \Gamma^{'}_{Z,Z} \, \nabla Z ).
\end{equation}

Considering that the progress variable $Y_c$ depends on $Z$ in the unburnt and burnt sides (denoted as $Y_{c,u}$ and $Y_{c,b}$, respectively), the normalized progress variable ($c$) can be defined as:

\begin{equation}\label{eq:normalization_Yc}
c(Y_c,Z)=\frac{Y_c-Y_{c,u}(Z)}{Y_{c,b}(Z)-Y_{c,u}(Z)},
\end{equation}

where the dependencies have been made explicit. Therefore, any quantity of the manifold $\psi$ is re-parametrized as $\psi=\psi(Y_c,Z) = \psi(c,Z)$. It is worth mentioning that this re-parametrization is only done for tabulation purposes even the transport equations to be solved still correspond to $Y_c$ and $Z$. This implies that the data is tabulated as a function of the $(c,Z)$ coordinates but the derivatives appearing in the coefficients $\Gamma^{'}_{Y_c,Y_c}$ and $\Gamma^{'}_{Z,Y_c}$ involve the evaluation of derivatives in the form $\partial / \partial Y_c |_Z$ and $\partial / \partial Z |_{Y_c}$, respectively (as usual the subscript after the vertical bar indicates the variable kept constant in the partial derivative).

Considering that $\psi=\psi(Y_c,Z) = \psi(c(Y_c,Z),Z)$, the derivatives required by the coefficients $\Gamma^{'}_{Y_c,Y_c}$, $\Gamma^{'}_{Z,Y_c}$, $\Gamma^{'}_{Y_c,Z}$ and $\Gamma^{'}_{Z,Z}$ can be obtained.
Including the dependency $\psi(c(Y_c,Z),Z)$ and the use of the definition (\ref{eq:normalization_Yc}), the different partial derivatives are expressed as:

\begin{equation}
\left . \frac{\partial \psi}{\partial Y_c}\right |_Z = \left . \frac{\partial \psi}{\partial c}\right |_Z \left . \frac{\partial c}{\partial Y_c}\right |_Z = \frac{1}{Y_{c,b}-Y_{c,u}} \left . \frac{\partial \psi}{\partial c}\right |_Z,
\end{equation}

\begin{equation}
\left . \frac{\partial \psi}{\partial Z}\right |_{Y_c} = \left . \frac{\partial \psi}{\partial c}\right |_Z \left . \frac{\partial c}{\partial Z}\right |_{Y_c} + \left . \frac{\partial \psi}{\partial Z}\right |_c,
\end{equation}

where

\begin{equation}\label{partial_c_partial_Z}
\left . \frac{\partial c}{\partial Z}\right |_{Y_c} = \frac{-1}{Y_{c,b}-Y_{c,u}} \left(\frac{d Y_{c,u}}{d Z} + c \, \frac{d }{d Z}(Y_{c,b} - Y_{c,u}) \right).
\end{equation}

In the following, it is assumed that the chemical manifold is parametrized by $(c,Z)$ and, therefore, it is assumed, unless otherwise stated, that $\partial / \partial c = \partial / \partial c |_Z$ and $\partial / \partial Z = \partial / \partial Z |_c$.

Starting with the transport equation for $Y_c$, to simplify the expressions and make them more compact, it is convenient to define the two terms that replicate the structure of the expression in Eq. (\ref{chain_rule_flux}), one associated to the scaled progress variable, $L_{Y_c,c}$, and the other to the mixture fraction, $L_{Y_c,Z}$:

\begin{equation}\label{L_Yc_c}
\begin{split}
L_{Y_c,c} = \sum_{k=1}^{N_s} & \alpha_k \Biggl(D_k \frac{\partial Y_k}{\partial c} +
\frac{D_kY_k}{W} \frac{\partial W}{\partial c} - \\
& -Y_k\sum_{j=1}^{N_s}D_j \frac{\partial Y_j}{\partial c} - 
  \frac{Y_k}{W} \frac{\partial W}{\partial c} \sum_{j=1}^{N_s}D_jY_j \Biggr) = \sum_{k=1}^{N_s} \alpha_k \Lambda_{Y_k,c},
\end{split}
\end{equation}

\begin{equation}
\label{L_Yc_Z}
\begin{split}
L_{Y_c,Z} = \sum_{k=1}^{N_s} & \alpha_k \Biggl(D_k \frac{\partial Y_k}{\partial Z} +
\frac{D_kY_k}{W} \frac{\partial W}{\partial Z} -  \\
& - Y_k\sum_{j=1}^{N_s}D_j \frac{\partial Y_j}{\partial Z} - 
 \frac{Y_k}{W} \frac{\partial W}{\partial Z} \sum_{j=1}^{N_s}D_jY_j \Biggr)=\sum_{k=1}^{N_s} \alpha_k \Lambda_{Y_k,Z}.
\end{split}
\end{equation}

The variables $L_{Z_p,c}$ and $L_{Z_p,Z}$ can be similarly defined for the mixture fraction of the element $p$:

\begin{equation}
L_{Z_p,c} = W_p \sum_{k=1}^{N_s}\frac{a_{kp}}{W_k}\,\Lambda_{Y_k,c},
\end{equation}

\begin{equation}
L_{Z_p,Z} = W_p \sum_{k=1}^{N_s}\frac{a_{kp}}{W_k} \, \Lambda_{Y_k,Z},
\end{equation}
which can be combined according to Eq. (\ref{eq_jZ}) to provide, in turn, the corresponding variables for $Z$, $L_{Z,c}$ and $L_{Z,Z}$:

\begin{equation}\label{L_Z_c}
L_{Z,c} = K_Z \left(\frac{L_{Z_C,c}}{mW_C} \,+ 
\, \frac{L_{Z_H,c}}{nW_H} \, - \, 2\frac{L_{Z_O,c}}{\nu_{\mathrm{O_2}}W_{\mathrm{O_2}}} \right),
\end{equation}

\begin{equation}\label{L_Z_Z}
L_{Z,Z} = K_Z \left(\frac{L_{Z_C,Z}}{mW_C} \,+ 
\, \frac{L_{Z_H,Z}}{nW_H} \, - \, 2\frac{L_{Z_O,Z}}{\nu_{\mathrm{O_2}}W_{\mathrm{O_2}}} \right) .
\end{equation}

The diffusion coefficients $\Gamma^{'}$ used in the final set of governing equations, Eqs.~(\ref{eq_solve_Yc}) and~(\ref{eq_solve_Z}), can now be computed as:

\begin{equation}\label{gamma_Yc_Yc}
\Gamma^{'}_{Y_c,Y_c} = \frac{L_{Y_c,c}}{Y_{c,b}-Y_{c,u}}  - D_{th},
\end{equation}

\begin{equation}\label{gamma_Yc_Z}
\Gamma^{'}_{Y_c,Z} =  L_{Y_c,c} \left . \frac{\partial c}{\partial Z}\right |_{Y_c} + L_{Y_c,Z},
\end{equation}

\begin{equation}\label{gamma_Z_Yc}
\Gamma^{'}_{Z,Y_c} = \frac{L_{Z,c}}{Y_{c,b}-Y_{c,u}},
\end{equation}

\begin{equation}\label{gamma_Z_Z}
\Gamma^{'}_{Z,Z} = L_{Z,c} \left . \frac{\partial c}{\partial Z}\right |_{Y_c} + L_{Z,Z} - D_{th},
\end{equation}
where $\left . \partial c/\partial Z\right |_{Y_c}$ is obtained from Eq. (\ref{partial_c_partial_Z}).

Finally, the algorithm for interpolation deserves some attention. There exist two major differences in 
composition space when moving from unity Lewis number conditions to detailed transport. First, derivatives in the phase space (associated with $\Gamma^{'}_{i,j}$ coefficients) must be computed and, second, the flame trajectories are curved. As indicated previously, the variables are tabulated in the $(c,Z)$ space. It is convenient, then, the use of a a regular mesh defined by the cartesian product of the vectors $c_i^*$ with $i=1,\ldots, n_c$ and $Z_i^*$ with $i=1,\ldots, n_Z$, that define the grids for the normalized progress variable and mixture fraction. 
An efficient and accurate interpolation algorithm is crucial since otherwise numerical noise may appear 
when computing the derivatives.

It is important to notice that, despite the advantages of working with $c$ instead of $Y_c$, the use of a normalized  progress variable usually introduces an extra-curvature in the surfaces for the thermochemical variables $\psi=\psi(c,Z)$. Hence, a direct interpolation in the $(c,Z,\psi)$ space tends to deteriorate the quality of the data and, consequently, the values for the $\Gamma^{'}_{i,j}$ coefficients.

This can be illustrated by the $Y_{\mathrm{H_2O}}$ surface. Taking the mass fraction of $Y_{\mathrm{H_2O}}$ 
as the progress variable in a hydrogen/air flame, it can be written, considering absence of water vapour in the unburnt gases:

\begin{equation}\label{eq:surf_YH2O}
Y_{\mathrm{H_2O}} = Y_c = c \, Y_{c,b}(Z).
\end{equation}

Then, $Y_{\mathrm{H_2O}}$ is a plane in the $(Y_c,Z,\psi)$ space. However, in the $(c,Z,\psi)$ space the $Y_{\mathrm{H_2O}}$ surface becomes blended by the contribution of $Y_{c,b}(Z)$ (specially around the stoichiometric mixture fraction). In general, transforming the surfaces from the $(Y_c,Z,\psi)$ space into the $(c,Z,\psi)$ introduces an extra-curvature in the surfaces that produce spurious under/overshoots when computing the $\Gamma^{'}_{ij}$ coefficients. It has been also checked that increasing the order of the interpolation, even reducing the interpolation error, still produces errors in the same order of magnitude.

\begin{figure*}[h]
    \centerline{\includegraphics[height=4.5cm]{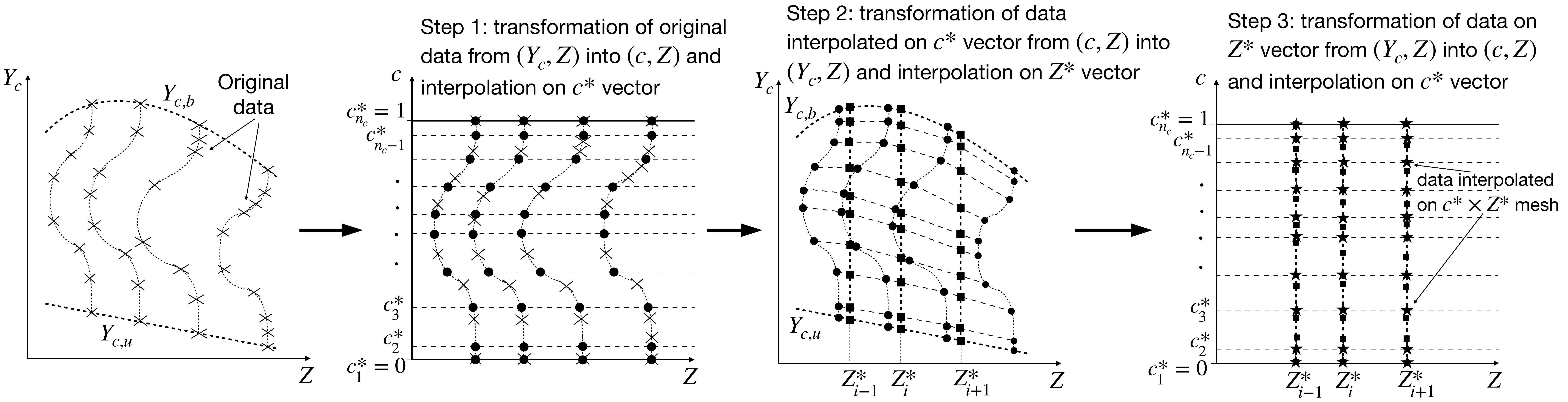}}
    \caption{Steps followed in the proposed interpolation algorithm.
}\label{fig:interpolation_steps}
\end{figure*}

Therefore, it is more convenient to carry out the interpolation in the $(Y_c,Z,\psi)$ space where curvatures are more moderate for the species mass fractions and enthalpy. The interpolation algorithm is illustrated in Fig. \ref{fig:interpolation_steps}. 

\begin{enumerate}
    \item In a first step, the dependent variables of the manifold are interpolated along the path described by each flame to obtain the data on the common vector $c^*$ (circle markers in Fig. \ref{fig:interpolation_steps}).
    \item Secondly, the interpolated data is taken to the $(Y_c,Z)$ space to find the corresponding values for each $Z_i^* \,\, i=1,\ldots, n_Z$. Since a linear interpolation is being used and $Y_{c,b}$ may show an important curvature (mainly around the stoichiometric mixture), the new interpolated data will not lay at the $c^*$ values. In this step both the dependent variables $\psi$ and $Y_c$ are interpolated (we refer to the interpolated values of $Y_c$ as $Y_{c,interp}$ which corresponds to the square markers in Fig. \ref{fig:interpolation_steps}) to find the new location of the interpolated in the $(Y_c,Z,\psi)$ space.
    \item Finally, the interpolated data, which lays on the $Z^*$ values, is interpolated into the $c^*$ mesh (star markers in Fig. \ref{fig:interpolation_steps}) from the knowledge of $Y_{c,interp}$ found previously. In this way, the values at the mesh defined by the cartesian mesh $c^* \times Z^*$ are obtained.
\end{enumerate}

Moreover, the regions of the domain out of the flammability limits are filled by extending the manifold to pure fuel and oxidizer. This extension becomes more relevant to incorporate the variations in the lean limit of the hydrogen flames due to preferential diffusion~\cite{Mizobuchi2019}.

The interpolation algorithm is applied to the species mass fractions $Y_i$ and enthalpy $h$ while the rest of the dependent variables of the manifold (temperature, diffusion coefficients, source terms, etc.) are obtained from the interpolated composition and enthalpy at the set of points defined by the mesh $c^* \times Z^*$.

The proposed algorithm for manifold representation is accurate and simple to use, since only one-dimensional interpolations are done at each step and, can be applied to multidimensional manifold representations. Moreover, as previously indicated, the use of higher order interpolation algorithms does not lead to a substantial reduction of the error when proceeding with the interpolation entirely in the $(c,Z,\psi)$ space.

\begin{figure*}[h]
    \centerline{\includegraphics{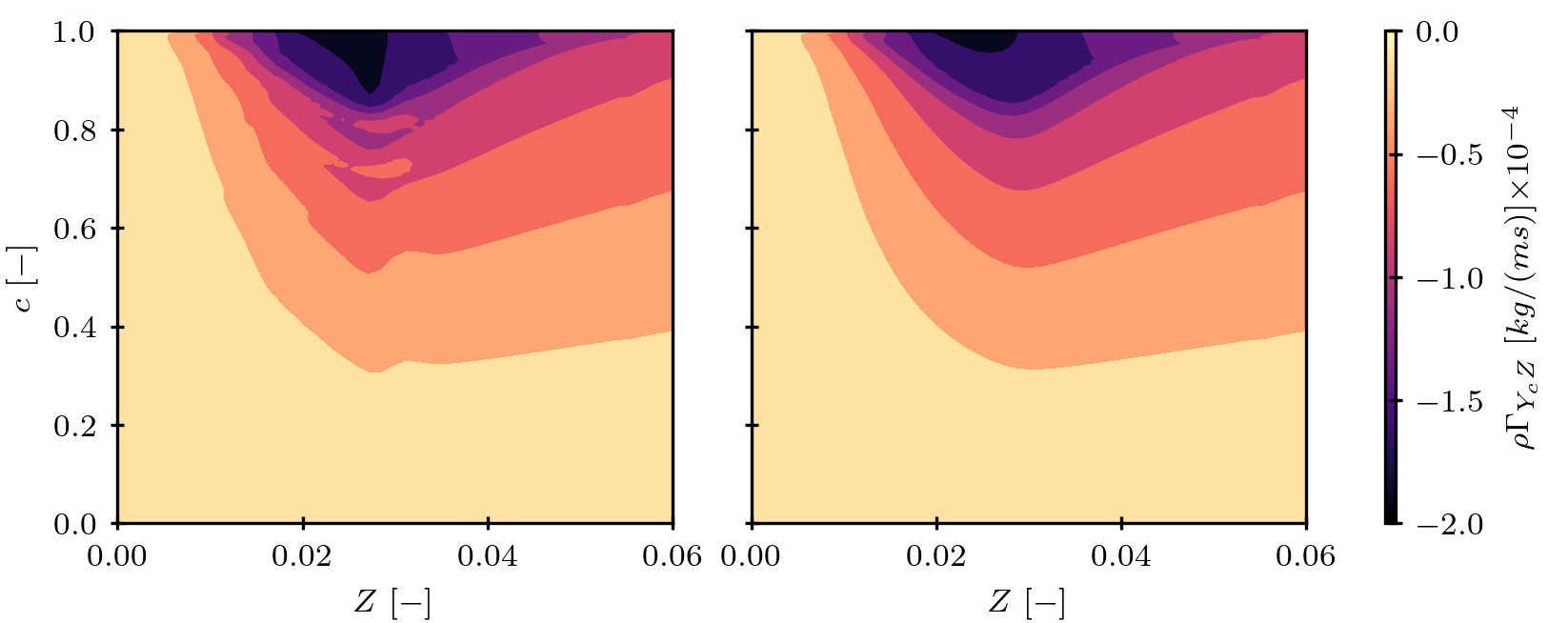}}
    \caption{Comparison of the coefficient $\rho \Gamma_{Y_c,Z}$ as a function of $c$ and $Z$ when interpolating entirely in the $(c,Z)$ (left) and applying previous interpolating algorithm (right). The discretization used for $c$ and $Z$ is given in section \ref{sec:results}.    }\label{fig:table_coefficients_different_interpolation}
\end{figure*}

Figure \ref{fig:table_coefficients_different_interpolation} shows the comparison of the tabulated 
$\rho \, \Gamma_{Y_c,Z}$ coefficients when interpolating entirely in the $(c,Z,\psi)$ space and when applying previous interpolation algorithm. Both maps are done with linear interpolations using the same meshes. The case corresponds to a planar premixed flame of hydrogen/air at atmospheric pressure and temperature $T=298.15$ K. The flames are computed from the lower flammability limit at $\phi=0.2$ up to $\phi=10$ extending the manifold as described for $\phi < 0.2$. Details related to the algorithm and discretization are given in sections \ref{sec:num_framework} and \ref{sec:results}. From Fig. \ref{fig:table_coefficients_different_interpolation} it is evident that the proposed interpolation algorithm produces more accurate results.

Finally, the full algorithm used to compute the $\Gamma^{'}_{ij}$ coefficients is summarized here:

\begin{enumerate}
\item Compute the 1D premixed laminar flames at a set of equivalence ratios.
\item Calculate $Y_{c,u}$ and $Y_{c,b}$ at each of the mixture fractions associated to previous flames.
\item Compute $(c,Z)$ coordinates for each point from $Y_{c,u}$ and $Y_{c,b}$ using Eq. (\ref{eq:normalization_Yc}).
\item Prescribe the boundaries of the manifold $Z=0$ and the maximum mixture fraction to be tabulated $Z=Z_{upper}$.
\item Interpolate the species mass fractions and enthalpy at the regular mesh generated by the cartesian products of vectors $c^*$ and $Z^*$ according to the previously described interpolation algorithm.
\item Compute the derived quantities (density, diffusion coefficients, etc.) at each node of the manifold.
\item Compute $\partial / \partial c$ and $\partial / \partial Z$ and calculate $L_{Y_c,c}$, $L_{Y_c,Z}$, $L_{Z,c}$ and $L_{Z,Z}$ according to Eqs. (\ref{L_Yc_c}), (\ref{L_Yc_Z}), (\ref{L_Z_c}) and (\ref{L_Z_Z}), respectively at each node.
\item Find the coefficients $\Gamma^{'}_{Y_c,Y_c}$, $\Gamma^{'}_{Y_c,Z}$, $\Gamma^{'}_{Z,Y_c}$ and $\Gamma^{'}_{Z,Z}$ according to Eqs. (\ref{gamma_Yc_Yc}) to (\ref{gamma_Z_Z}) respectively.
\end{enumerate}

\section{Numerical framework}\label{sec:num_framework}

\subsection{Governing equations}

The results obtained by the flamelet model with mixture-averaged transport (FGM) will be compared with those obtained by solving the governing equations of species mass fractions with a detailed chemistry model and mixture-averaged diffusion model, referred here as DC. A numerical framework
including both approaches have been developed in the multiphysics code Alya~\cite{Vazquez2016AlyaMultiphysicsEngineering} from the Barcelona Supercomputing Center using the same numerical methods and discretization strategies. Equations are solved using a low Mach number approximation of the Navier-Stokes equations~\cite{Both_CAF_2020}, for which the continuity and momentum equations are given by:

\begin{equation}
    \label{eq:NS_cont}
    \frac{\partial \rho}{\partial t} + \nabla \cdot \left( {\rho} \boldsymbol{u} \right) = 0
\end{equation}
    
\begin{equation}
    \label{eq:NS_mom}
    \frac{\partial (\rho \boldsymbol{u})}{\partial t} + \nabla \cdot (\rho \boldsymbol{u} \otimes \boldsymbol{u}) = -\nabla p + \nabla \cdot \boldsymbol{\tau},
\end{equation}

where $p$ and $\boldsymbol{\tau}$ represent the pressure and shear stresses tensor, respectively.

For the DC model, the evolution of the multicomponent gas is expressed in terms of the transport equations for each of the $N_s$ individual species $Y_k$, Eq. (\ref{eq:transport_Yk}), which is replicated here for completeness:

\begin{equation}
    \label{eq:Yk}
    \rho \frac{\partial Y_k}{\partial t} + \rho \boldsymbol{u} \nabla \cdot Y_k + \nabla  \cdot \mathbf{j}_k = \rho \, \dot\omega_{k},
\end{equation}

where $\mathbf{j}_k$ is the diffusion flux given by mixture-averaged model, see Eq.~(\ref{diff_eq_MA_total}). In Eq.~(\ref{eq:Yk}), $\dot\omega_{k}$, which denotes the chemical source term for species $Y_k$, is given by the Arrhenius law and the law of mass action. For the enthalpy, Eq.~(\ref{eq_h}) is solved:

\begin{equation}\label{eq_h2}
\rho \frac{\partial h}{\partial t} + \rho \boldsymbol{u} \cdot \nabla h + \nabla \cdot (\rho \, \mathbf{j_h} )= 0,
\end{equation}

with the diffusive flux $\mathbf{j_h}$ given by the Eq.(\ref{eq_jh}). Finally, the ideal gas state equation is applied. The DC model in Alya has been applied in previous studies~\cite{MIRA2022,RAMIREZMIRANDA2023105723,Surapaneni2023AssessmentDynamicAdaptive, KALBHOR2023112868}.

Regarding the FGM model, the continuity and Navier-Stokes equations are solved together with Eqs. \eqref{eq_solve_Yc} and \eqref{eq_solve_Z} for the progress variable and mixture fraction, respectively:

\begin{equation}\label{eq_solve_Yc2}
\rho \frac{\partial Y_c}{\partial t} + \rho \boldsymbol{u} \cdot \nabla Y_c = \nabla \cdot (\rho \, D_{th} \, \nabla Y_c +\rho \, \Gamma^{'}_{Y_c,Y_c} \, \nabla Y_c  + \rho \, \Gamma^{'}_{Y_c,Z} \, \nabla Z ) +
\rho \, \dot{\omega}_{Y_c},
\end{equation}

\begin{equation}\label{eq_solve_Z2}
\rho \frac{\partial Z}{\partial t} + \rho \boldsymbol{u} \cdot \nabla Z = \nabla \cdot (\rho \, D_{th} \, \nabla Z +\rho \, \Gamma^{'}_{Z,Y_c} \, \nabla Y_c  + \rho \, \Gamma^{'}_{Z,Z} \, \nabla Z ).
\end{equation}

The flamelet model implementation in Alya with no preferential diffusion effects has been used in previous works \cite{Govert2015,Govert2018,BENAJES2022111730,mira2018numerical,mira2020numerical}.

\subsection{Numerical discretization}

Simulations have been carried out with the parallel multi-physics code Alya \cite{Vazquez2016AlyaMultiphysicsEngineering}
developed at the Barcelona Supercomputing Center. Alya solves the low-Mach number approximation of the Navier-Stokes equations using a low-dissipation single stage numerical scheme for the velocity and pressure coupling~\cite{Both_CAF_2020}. The scalar transport involving the enthalpy, species transport equations and control variables for the flamelet model, are stabilized using the Algebraic Subgrid Scale (ASGS) method \cite{Castillo2014}. The spatial discretization for momentum and scalars is based on the Finite Element Method with a second order scheme using linear elements. All governing equations use an explicit third-order Runge-Kutta scheme to advance in time. The flamelet solutions for the tabulation are obtained by solving unstretched adiabatic premixed flames using the Cantera solver (version 2.6.0)~\cite{cantera}. Besides, a skeletal mechanism containing 9 species and 12 reactions for hydrogen-air combustion from Boivin et al. \cite{Boivin2011ExplicitReducedMechanism} is used for all the calculations in Alya and Cantera.

\section{Results and discussion}\label{sec:results}

In this section, several flame configurations are defined
in order to examine the capability of the proposed model to recover the burning rates of hydrogen flames when including detailed transport phenomena. Before considering spatial variations in reactant composition, an analysis of one-dimensional premixed flames at different equivalence ratios is presented in subsection~\ref{sec:1D}. The solutions of the model are compared with those obtained by Cantera and the model is tested from lean to rich conditions. Different levels of stratification are considered in the next subsections in order to identify the limits of the model to account for two-dimensional effects across the flame front including preferential diffusion. 
In subsection~\ref{sec:2D-GradZ}, a configuration based on a flame front interacting with a mixture fraction gradient is analysed. Different ratios of mixing length over flame thicknesses are considered to characterize the fluxes of $Z$ and $Y_c$ across the flame. Finally, subsection~\ref{sec:2D-TripleFlame} deals with the problem of a triple flame, where the mixture fraction variation expands over the flammability range, and these two-dimensional effects become stronger. These cases have been chosen as they correspond to canonical configurations whose modelling is not only challenging, but also play a major role in the flame propagation and stabilization of partially premixed mixtures as those found in lifted jet flames \cite{Muller1994,Ray2000}, mixing layers and auto-ignition fronts \cite{Chung2007,Mastorakos2009}.

For all the cases presented here, mixture-averaged transport is included in the finite rate chemistry and flamelet models, and heat loss effects are neglected to keep the dependency of the thermal states on the progress variable and mixture fraction  only ($\psi=(Y_c,Z)$). The thermochemical conditions for all the cases presented in the following correspond to hydrogen-air mixtures at atmospheric pressure and a temperature for the unburnt gases of $T=298.15$ K.

\begin{figure*}[h]
    \centerline{\includegraphics[width=15.0cm]{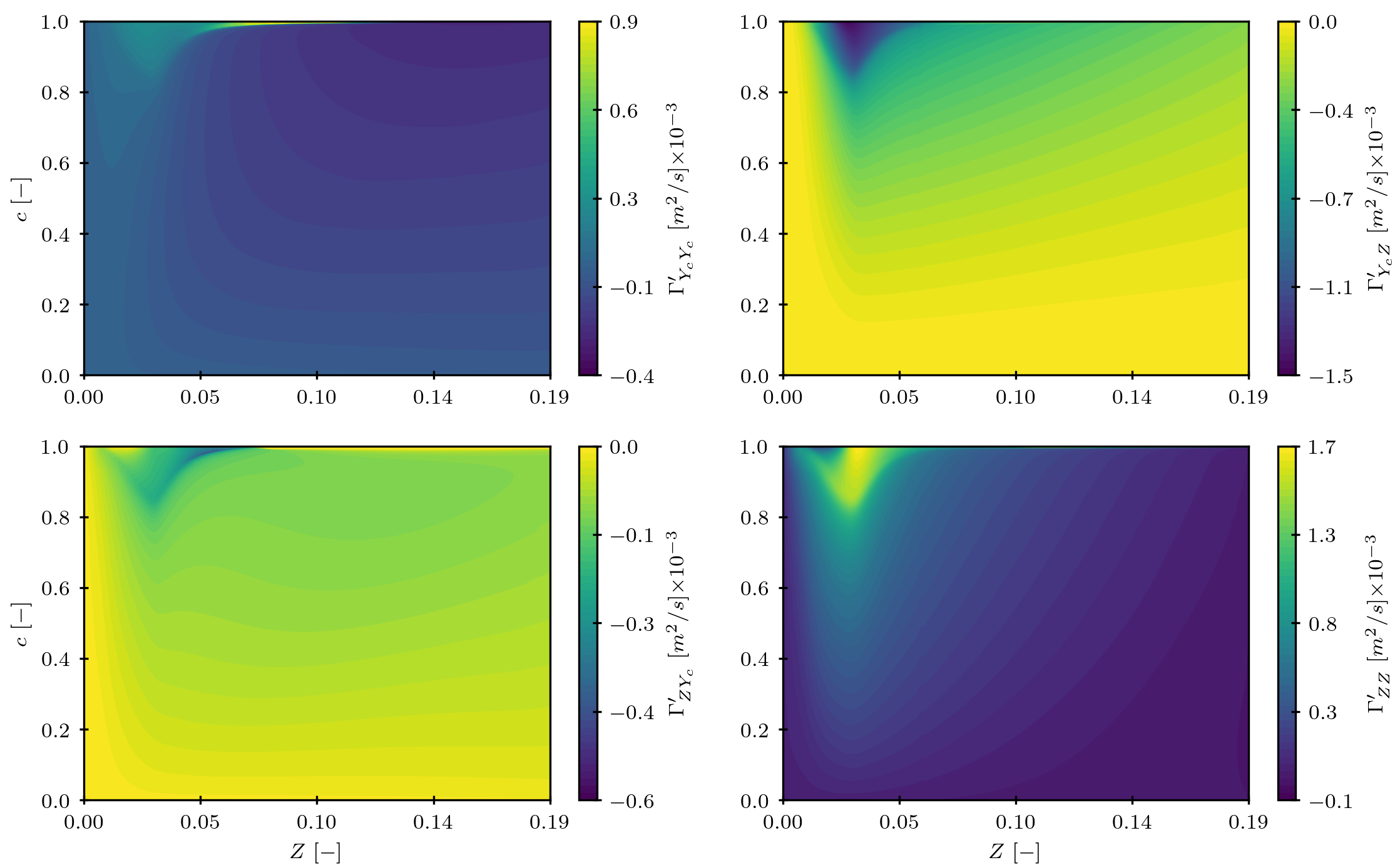}}
    \caption{$\Gamma^{'}_{Y_c,Y_c}$, $\Gamma^{'}_{Y_c,Z}$, $\Gamma^{'}_{Z,Y_c}$ and $\Gamma^{'}_{Z,Z}$ coefficients in the $(c,Z)$ space computed from one-dimensional adiabatic hydrogen flames at atmospheric pressure and unburnt gas temperature of 298.15 K.
    }\label{fig:table_coefficients_colormap}
\end{figure*}

The flamelet manifold used for the problems presented below was generated using two control variables: the progress variable defined by $\mathcal{Y}_{1} = Y_{\mathrm{H_2O}}$ and the Bilger's mixture fraction $\mathcal{Y}_{2} = Z$. Even this definition for the progress variable is injective in the flammability region, it is worth mentioning that its descriptive capability is deteriorated for rich mixture fractions near equilibrium as it shows strong gradients in these regions. A total of 96 flames have been computed to create the flamelet manifold. A mixture fraction distribution with non-uniform discretization and refinement around the stoichiometric mixture was chosen. The progress variable $c$ space is defined by a total of 81 points using an arithmetic progression between 0 and 0.9 with uniform spacing $\Delta c=0.01$ and a much finer resolution with $\Delta c=0.001$ near equilibrium (from 0.9 to 1) due to the aforementioned degradation of the progress variable close to the burnt conditions for the rich mixtures. The resulting maps for $\Gamma^{'}_{Y_c,Y_c}$, $\Gamma^{'}_{Y_c,Z}$, $\Gamma^{'}_{Z,Y_c}$ and $\Gamma^{'}_{Z,Z}$ in the $(c,Z)$ space for the set of flames at the conditions described previously are
shown in Fig.~\ref{fig:table_coefficients_colormap}.

\subsection{One-dimensional premixed flames}\label{sec:1D}

In a first step, a validation of the model in a one-dimensional adiabatic premixed flame configuration is carried out. One-dimensional flame solutions from Cantera, which are used to build the database, are compared with the results obtained from the solution of the FGM model. Note that this flame configuration could be solved from a one-dimensional manifold with $Y_c$ as a the only control variable. However, using a two-dimensional manifold based on $Z$ and $Y_c$ provides a more consistent validation of the method, since the flame has more degrees of freedom and there is no mechanism that forces the flame to follow the flame trajectory given by the DC model. Therefore, any deviation in flame trajectory by numerical errors or inaccuracies of the model will cause 
noticeable deviations in the flame structure. This validation case is set to ensure that the model recovers the flame speed and thickness as well as the flame structure of the reference flame when including mixture-averaged transport.

To solve the one-dimensional equations, a mesh composed of linear elements with a uniform spacing defined by the parameter $n_x$ is used such that the spacing between elements follows $\Delta x = l_F / n_x$, where $l_F$ is the flame thickness defined in the following, (see Eq.~(\ref{eq:thermal_flame_thickness})). For the results shown here, $n_{x}=20$ is selected as it reproduces well the reference results from Cantera. The boundary conditions for the gas phase are set with Dirichlet condition for the mixture composition at different equivalence ratios 
at the inlet and Neumann conditions at the outlet. To proceed with the validation, an assessment of the chemical structure and macroscopic properties of the flame is performed. The laminar flame speed $s_L$ \cite{Poinsot2011TheoreticalNumericalCombustiona} is one of the most relevant parameters of a premixed flame and is highly sensitive to the transport model, specially in hydrogen flames. For one-dimensional flames it can be indistinctly defined as:

\begin{equation}\label{eq:laminar_flame_speed}
s_L = \frac{u_b - u_u}{\rho_u / \rho_b - 1} = -\frac{1}{\rho_u Y_{F,u}} \int_{-\infty}^{+\infty} \dot{\omega}_F d x,
\end{equation}

where the sub-indexes $u$ and $b$ refer to the unburnt and burnt mixture state, respectively, and $Y_F$ and
$\dot{\omega}_F$ denote the fuel mass fraction and the fuel chemical source term, respectively. Besides the flame speed, the premixed flame is characterized by a given thickness, which is influenced by the diffusive characteristics of the front. The flame thickness $l_{F}$ can be estimated by the maximum temperature gradient and the temperature difference between the burnt and unburnt conditions~\cite{Poinsot2011TheoreticalNumericalCombustiona}.

\begin{equation}\label{eq:thermal_flame_thickness}
    l_{F} = \frac{T_{b} - T_{u}}{\mathrm{max}(|\nabla T|)},
\end{equation}
where $T_{u}$ and $T_{b}$ are the unburnt and burnt temperatures, respectively. This quantity is used in this work to define a length scale for the flame and a characteristic time scale $\tau = l_{F} / s_{L}$.

The comparisons of laminar flame speed $s_L$ and thermal flame thickness $l_F$ between Cantera and the proposed model are presented in ~\ref{appendix:1D_sub1}. The results for the flame speed have a consistent relative error with a maximum value of $\epsilon = | (s_L^{FGM} - s_L^{C}) \, / \, s_L^{C} |=0.85\%$, while a maximum relative error of $\epsilon = | (l_F^{FGM} - l_F^{C}) \, / \, l_F^{C} |=2.04\%$ is obtained for the flame thickness  
(superscript $C$ refers to calculations with Cantera) in the range $\phi \in [0.6,6.5]$. Only for $\phi=0.5$ these errors are magnified until 1.56\% and 2.35\% for the flame speed and thickness, respectively, even such errors are still considered acceptable.

\begin{figure*}[h]
    \centering
    \centerline{\includegraphics{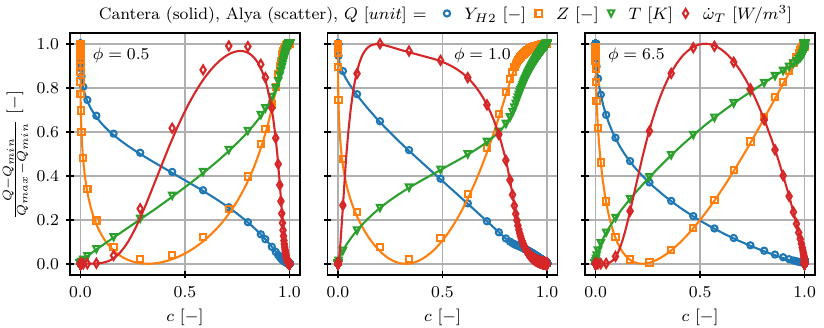}}
    \caption{Normalized scalar fields for $Y_{\mathrm{H_2}}, Z, T$ and $\dot{\omega}_T$ (heat release) coloured by blue, orange, green and red, respectively, for Cantera reference solution (solid lines) and Alya with FGM and mixture-averaged transport (markers) as a function of the normalized progress variable along the flames for the equivalence ratios of $0.5$ (left), $1.0$ (center) and $6.5$ (right). Results for one-dimensional adiabatic hydrogen/air flames at atmospheric pressure and unburnt gas temperature of 298.15 K.} \label{fig:alya_vs_ct}
\end{figure*}

To analyse the thermochemical structure of the flame, profiles of the relevant quantities $Y_{\mathrm{H_2}}$, $Z$, $T$ and $\dot{\omega}_T$ are represented in progress variable space $c$ (see Eq. (\ref{eq:normalization_Yc})) 
in Fig.~\ref{fig:alya_vs_ct} for the given set of equivalence ratios $\phi \in \left\{ 0.5, 1.0, 6.5 \right\}$. These values correspond to relevant points in the manifold space: values of $\phi=0.5$ and $\phi=6.5$ correspond to mixtures close to the flammability limits, which will play a key role in the triple flame configuration analysed in section \ref{sec:2D-TripleFlame}. The results show an excellent prediction of all the quantities from lean to rich conditions and demonstrate the ability of the model to recover the flame structure with preferential diffusion.

To complete this section, an analysis of the contribution of the terms arising from the diffusion transport in Eqs. \ref{eq_solve_Yc2} and \ref{eq_solve_Z2} through the study of $\Gamma^{'}_{i,j}$ coefficients is given in \ref{appendix:1D_coeffs}.

These results validate the ability of the proposed modelling to accurately describe the flame structure and characteristics when preferential diffusion is present. In the following, two configurations with different levels of stratification of the fresh mixture are considered. The first one, with a lower level of complexity, corresponds to a stratified flame with limited variation in composition. Therefore, in this case, even the gradients in mixture fraction can be high, the difference between the minimum and maximum mixture fraction is limited to a small value. On the contrary, the second configuration corresponds to a triple flame where the range of mixture fractions encompass the flammability range and a more complex flame structure is formed.

\subsection{Stratified flame with limited variation in composition}\label{sec:2D-GradZ}

This section presents the problem of a flame front interacting with reactants with a limited variation in composition. The inlet condition is defined by an inlet mixture fraction with a spatial profile so that the flame exhibits lean and rich fronts with periodic conditions. The configuration has been selected to asses the model under conditions involving mixture fraction gradients with a flamelet-like structure without the presence of a diffusion flame, as it occurs in the triple flame configuration presented in the next subsection.

\begin{figure}
    \centerline{\includegraphics{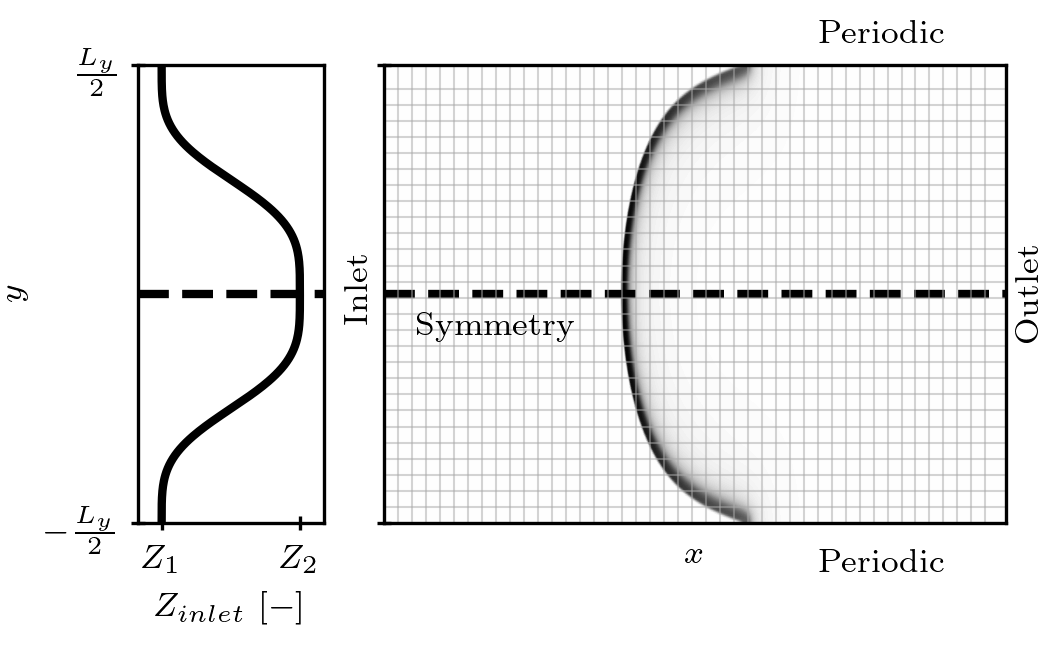}}
    \caption{
    Sketch of the domain and flame front indicating the injection profile of mixture fraction $Z_{in}\left(y\right)$ from Eq. \eqref{eq:2D-GradZ_inlet}. The square elements are included for illustrative purposes and do not correspond to the real mesh.}
    \label{fig:2D-GradZ_sketch}
\end{figure}

The configuration consists of a rectangular domain of length $L_x$ and height $L_y$ with a mixture fraction profile $Z(0,y) = Z_{in}(y)$. Figure ~\ref{fig:2D-GradZ_sketch} shows the computational domain and flame configuration. For the mesh discretization, square elements of length $\delta_{x} = l_{F}\left(\phi=1.0\right)/10$ are set.

The mixture fraction profile of the inlet function $Z_{in}$ is given by:

\begin{equation}
\label{eq:2D-GradZ_inlet}
Z_{\text{in}}=\frac{Z_{1}+Z_{2}}{2}+\frac{Z_{1}-Z_{2}}{2} \cos \left\{\frac{\pi}{2}\left[1+\cos \left(\frac{2 \pi y}{L_y}\right)\right]\right\} \quad \quad y \in \left[-\frac{L_y}{2},\frac{L_y}{2} \right]\text {. }
\end{equation}

As the length $L_y$ of the mesh determines the magnitude of the gradients, the mixing length $\delta_{m}$ at the inlet is used to define four flame configurations (I to IV), see Table~\ref{tab:2D-GradZ_table}. For all the simulations $L_x$ is kept equal to a value of 25 mm. The mixing length $\delta_{m}$ is defined by:

\begin{equation}
\label{eq:2D-GradZ_inlet-mixing-length}
\delta_{\mathrm{m}}=(Z_2-Z_1)\left(\left.\frac{d Z_{in}}{d y}\right|_{y=L_y / 4}\right)^{-1}.
\end{equation}

\begin{table}
    \centering
    \captionof{table}{Dimensions of the domain and mixing length for each of the four simulated cases. The cases are ordered by increasing mixing length.\label{tab:2D-GradZ_table}}
    \begin{tabular}{|c|c|c|}
    \hline
    Index & $\delta_m$/$l_F(\phi=1)$ & $L_y /2$ (mm) \\ \hline
    I &  1.03  & 0.85    \\ \hline
    II & 2.11  &  1.75   \\ \hline
    III & 3.02 & 2.5  \\ \hline
    IV & 4.84  & 4 \\ \hline
    \end{tabular}
\end{table}

The boundary conditions for the simulation set-up are as follows: for all the cases, a mixture fraction profile 
ranging from $Z_1=Z(\phi=0.8)\approx0.023$ to $Z_2=Z(\phi=1.2)\approx0.034$ is defined at the inlet and is used to define the species mass fractions and enthalpy for the DC model. Periodic boundary conditions are imposed at the bottom and top sides of the domain, and zero Neumann conditions are prescribed at the outlet. The initial condition is chosen as $Z(x,y)|_{t=0}=Z_{in}(y)$ for the mixture for all the domains, coupled with a step function from $c=0$ to $c=1$ at $x_{ic}=1/5L_{x}$ for flame I, $x_{ic}=2/5L_{x}$ for flame II and $x_{ic}=3/5L_{x}$ for flames III and IV. As $L_{y}$ increases, the flame extends along a larger part of the physical space, forcing to adjust the initial condition to prevent the flame from getting too close to the inlet.

The variation of $Z$ from $Z_1$ to $Z_2$ across the inlet function leads to a monotonic increase in laminar flame speed from $s_L(Z_1)$ to $s_L(Z_2)$ (recall that $s_L$ is maximum at $\phi=1.6$ approx.). Because of the absence of heat losses in this configuration, its structure can be determined solely by $Z$ and $Y_{c}$. However, when the mixing length (defined by Eq. (\ref{eq:2D-GradZ_inlet-mixing-length})) is relatively small, it is expected the flame curvature to get higher and, hence, to have an increase of the tangential flux of diffusion with respect to to the normal flux \cite{VanOijen2004NumericalStudyConfined}. The two-dimensional structure of this configuration can have, therefore, an effect on the predictions of the FGM since the manifold is constructed based on one-dimensional flames. Even it can be argued that for many fuels the flames show chemical reactions layers much smaller than the flame thickness and, hence, the effect of the curvature and stretch is limited on the prediction of flame propagation speed and shape \cite{VanOijen2004NumericalStudyConfined}, in the case of hydrogen, the modelling is more challenging. The high diffusivity of the fuel spreads the reaction layer over an important fraction of the flame thickness (see Fig. \ref{fig:alya_vs_ct}) as can be verified from the distribution of heat release rate in $c$ space.

\begin{figure}
    \centerline{\includegraphics[width=15.0cm]{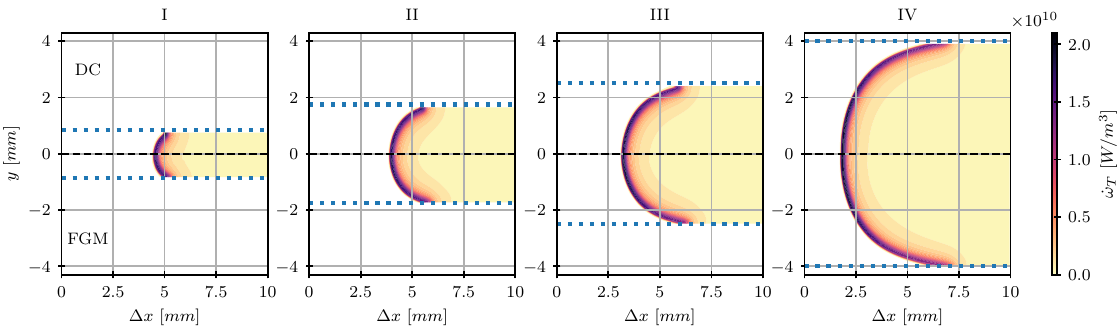}}
    \caption{Contour plots of $\dot{\omega}_{T}$ for the simulation setup of \ref{sec:2D-GradZ} for cases from Table \ref{tab:2D-GradZ_table}, taken after the transient regime at time $t=5.0$ ms. Top half corresponds to DC solution while bottom half to FGM. Blue dots indicate the size of the domain in the transversal direction to the flow. All flames are centered for visualization.}
    \label{fig:2D-GradZ_ColormapMixingLengthTransition}
\end{figure}

The effect of the mixing length transition from high to low gradients can be distinguished in Fig.~\ref{fig:2D-GradZ_ColormapMixingLengthTransition}, which shows contour plots for the heat release, defined by $\dot{\omega}_{T}= \sum_{k=1}^{N_{s}} \Delta h_{f,k}^0 \dot{\omega}_{k}$, where $\Delta h_{f,k}^0$ is the formation enthalpy for the k-th species at the ambient temperature for both DC and FGM. On the one hand, when the mixing length is large, diffusion is less intense and cannot alter the inlet reactant composition leading to a flame with low gradients in composition. Therefore, the flame exhibits a significant disparity of flame speeds (see Fig. \ref{fig:flame_speed}), but the tangential fluxes are expected to have a limited relevance (case IV). On the other hand, it is observed that when the mixing length is small (high mixture fraction gradients), the strong diffusive fluxes appearing upstream of the flame tend to rapidly reduce the jump in mixture fraction and, hence, flame speed variability. However, the gradients in composition are high leading to a curved flame front, as observed in cases I and II, where the tangential fluxes may become more important. Even in this case, the flame shape and the leading edge dynamics are well recovered by the tabulated approach as observed in Fig. \ref{fig:2D-GradZ_ColormapMixingLengthTransition}. A more detailed analysis for the comparison of DC and FGM is given in the following.

\begin{figure*}[h]
    \centerline{\includegraphics{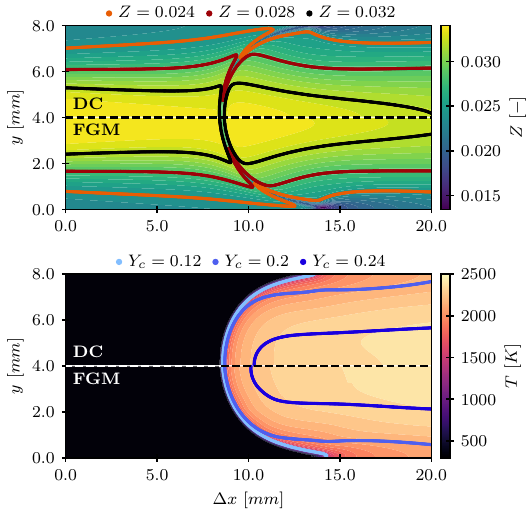}}
    \caption{Contour plot of $Z$ (top) and $T$ (bottom) for the simulation setup of \ref{sec:2D-GradZ} for the flame IV from Table \ref{tab:2D-GradZ_table} at time $t=5$ ms. The area of $y>4mm$ corresponds to DC while bottom to FGM results. Additionally, the iso-lines for $Z$ isolines associated to the values $Z=0.024$ (light orange), 0.028 (orange), 0.032 (dark red) as well as the iso-lines for $Y_{c}$ for values $Y_{c}=0.12$ (light blue), 0.2 (blue) and 0.24 (dark blue) are included.}\label{fig:2D-GradZ_ColormapMultiple}
\end{figure*}

\begin{figure*}[h]
    \centering
    \centerline{\includegraphics{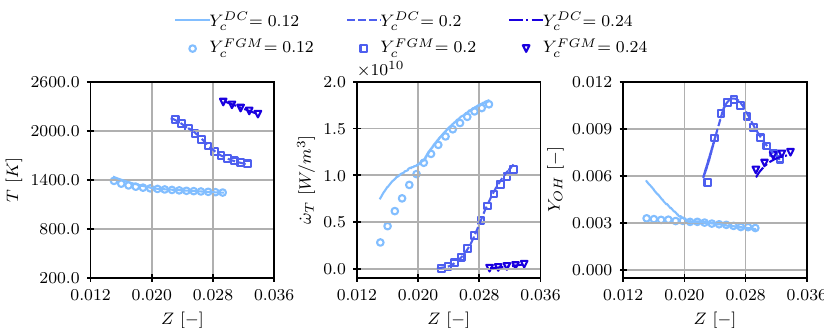}}
    \caption{Values of $T$, $\dot{\omega}_T$ and $Y_{\mathrm{OH}}$ vs $Z$ along the isolines of values $Y_{c}=0.12$ (light blue), $0.2$ (blue) and $0.24$ (dark blue) for DC model (lines) and FGM (empty markers) for flame IV. These isolines are shown in the temperature map from Fig. \ref{fig:2D-GradZ_ColormapMultiple}.}\label{fig:2D-GradZ_isoYc}
\end{figure*}

\begin{figure*}[h]
    \centering
    \centerline{\includegraphics{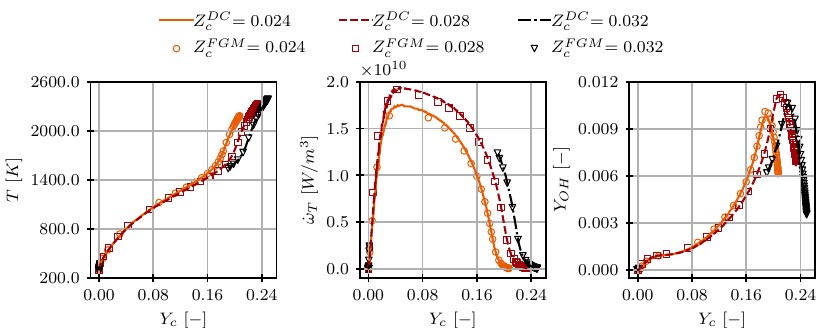}}
    \caption{Values of $T$, $\dot{\omega}_T$ and $Y_{\mathrm{OH}}$ vs $Y_{c}$ along isolines of values $Z=0.025$ (light orange), $0.028$ (orange) and $0.032$ (dark red) for DC model (lines) and FGM (empty markers) for flame IV. These isolines are shown in the mixture fraction map from Fig. \ref{fig:2D-GradZ_ColormapMultiple}. }\label{fig:2D-GradZ_isoZ}
\end{figure*}

The results of the flame configuration IV (lowest $Z$ gradient) are shown in Fig.~\ref{fig:2D-GradZ_ColormapMultiple} by the contour plot of $Z$ and $T$, this last one downstream the flame front region. Due to the symmetry of the configuration, maps for the DC and FGM models are presented on the top and bottom halves of the plot, respectively. Moreover, iso-lines for mixture fraction $Z=0.024$, $0.028$ and $0.032$ as well as progress variable $Y_{c}=0.1$, $0.2$ and $0.24$ are included for reference. The results show a notable agreement between the models with only small discrepancies in the burnt gas side where the mixture fraction shows slightly smaller gradients with FGM. A more quantitative analysis of the flame structure for flame IV is shown in Figs.~\ref{fig:2D-GradZ_isoYc} and~\ref{fig:2D-GradZ_isoZ}. Profiles for $T$, $\dot{\omega}_{T}$ and $Y_{\mathrm{OH}}$ are represented along the level curves of progress variable $Y_c$ and mixture fraction $Z$ included in Fig. \ref{fig:2D-GradZ_ColormapMultiple}.

Overall, the results correctly redict the different quantities across the iso-lines demonstrating that the FGM can capture the internal structure of the stratified flame. In the case of the iso-line for $Y_c=0.12$ some deviations are observed for $Z<0.02$ (see Fig. \ref{fig:2D-GradZ_ColormapMultiple}). Such lean regions are found very close to the transversal limits of the domain (recall that the minimum $Z$ at the inlet is 0.023) where the iso-lines tend to become parallel to the domain and can be slightly affected by the boundary conditions. Moreover, it is observed in Fig. \ref{fig:2D-GradZ_isoZ} that the iso-line for $Z=0.032$ does not extend along the entire range of $Y_c$. This is also observed in Fig. \ref{fig:2D-GradZ_ColormapMultiple} where it is verified that such iso-line does not cross the flame front due to preferential diffusion effects. Therefore, even in this case, the model captures this aspect accurately showing an excellent agreement with the DC model. 

\begin{figure*}[h]
    \centerline{\includegraphics[width=15.0cm]{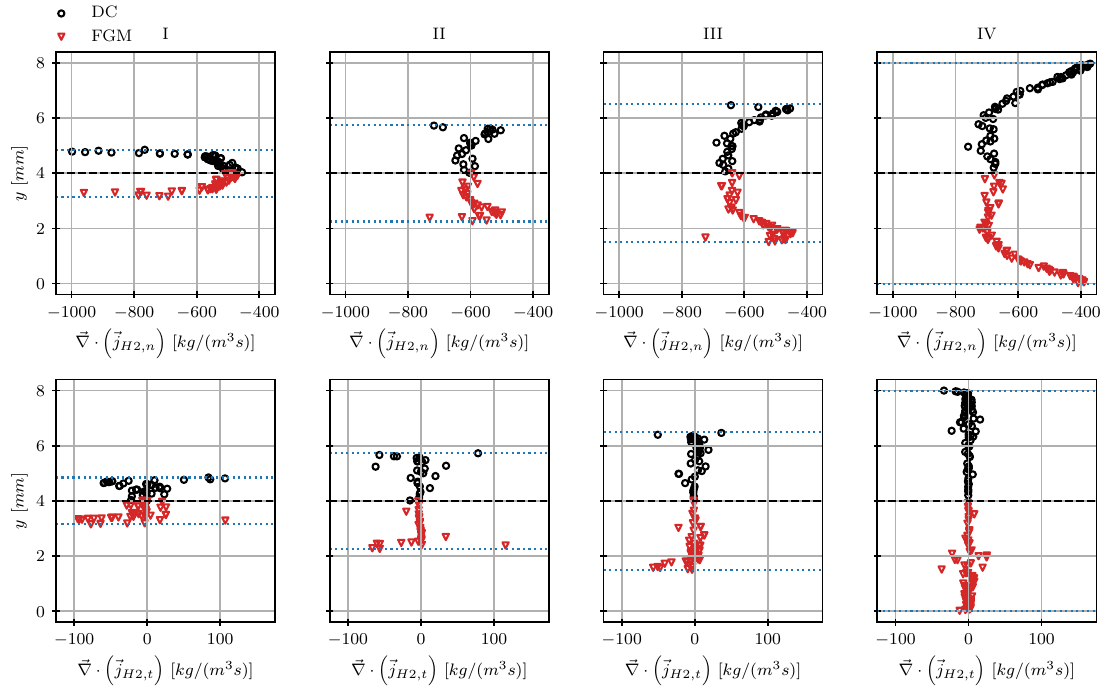}}
    \caption{Values for the normal (top subplots) and tangential (bottom subplots) diffusive flux balances for $\mathrm{H_2}$ for all the flames from Table \ref{tab:2D-GradZ_table}. All plots are centred around the middle point of flame IV, and blue dotted lines delimit their physical domains. For each subplot, the values for the DC model are shown in black circles from the line $y=4mm$ upwards, and the FGM as red triangles at the bottom of $y=4mm$. Notice that the scales of the figures are adapted to the real range of values of the variables.}\label{fig:2D-GradZ_DiffusiveFlux}
\end{figure*}

To properly quantify the effect of the bidimensionality of the problem on the predictive capabilities of the FGM model, the diffusive flux balance for the fuel $\nabla \cdot (\mathbf{j_{\mathrm{H_2}}})$ is decomposed into its normal and tangential components. Figure~\ref{fig:2D-GradZ_DiffusiveFlux} shows such decomposition across the flame front along the iso-line of $T=1000$ K. Then, the flux balances are defined as $\nabla \cdot (\mathbf{j_{\mathrm{H_2},n}})$ and $\nabla \cdot (\mathbf{j_{\mathrm{H_2},t}})$ where $\mathbf{j_{\mathrm{H_2},n}} = (\mathbf{j_{\mathrm{H_2}}} \cdot \mathbf{n}) \mathbf{n}$ for the normal component and  $\mathbf{j_{\mathrm{H_2},t}} = (\mathbf{j_{\mathrm{H_2}}} \cdot \mathbf{t}) \mathbf{t}$ for the tangential one. $\mathbf{n}$ and $\mathbf{t}$ denote the unitary normal and tangential vectors to the temperature iso-line, respectively. On the one hand, it is observed that all the flames exhibit values of the normal component flux balance in the same order of magnitude along the centerline even there is a reduction (in absolute value) when decreasing the mixing length. However, strong variations are observed towards the lateral boundaries, which are caused by the variations in flame shape for the different cases. On the other hand, the tangential contribution is negligible for high mixing lengths while it becomes more relevant when reducing the mixing length, first at the sides of the flame and later, for even smaller mixing lengths, at the center. Therefore, the role of the tangential component is more relevant when decreasing the mixing length as expected. However, in all the cases it is observed that the FGM model is able to reproduce the DC results with great accuracy only deviating from the reference very close to the boundaries.

Besides the flame structure, it is important to evaluate the ability of the FGM to predict the flame propagation speed and local variations of the burning velocity due to gradients in reactant composition. To evaluate the burning velocity, the consumption speed $s_C$ defined by the integral of the fuel sink term is used. In fully premixed conditions, the consumption speed can be obtained by integrating such term in the domain according to $s_{C} = (\rho_u Y_{F,u} L_{y})^{-1}\int_{S}\, \dot{\omega}_F\,dS$, where the subscript $u$ refers to the unburnt state. Because of the stratification, the previous formula is modified to account for the variations in fuel composition at the inlet:

\begin{equation}\label{eq:fuel_consumption}
s_{C} = \frac{ \int_{S}\, \dot{\omega}_F\,dS}{ \int_{0}^{Ly}\, \rho_u(0,y) Y_{F,u}(0,y)\,dy}.
\end{equation}

\begin{figure*}[h]
    \centerline{\includegraphics{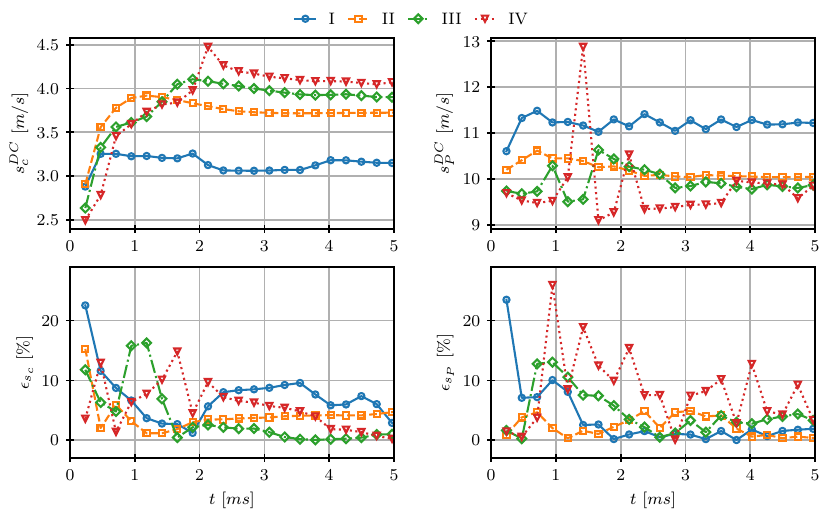}}
    \caption{Speed comparison for all the flames from table \ref{tab:2D-GradZ_table}. Top left: consumption speed $s_{C}$ for DC model. Top right: relative velocity $u_{rel,P}$ for point P for DC model. Bottom left: relative difference $\epsilon_{s_C} = |(s_C^{FGM} - s_C^{DC}) / s_C^{DC}|$ between models. Bottom right: relative difference $\epsilon_{u_{rel,P}} = | (s_P^{FGM} - s_P^{DC}) / s_P^{DC} |$ between models.}\label{fig:2D-GradZ_Speed}
\end{figure*}

While the consumption speed accounts for the flame propagation using the local burning rates, it is also relevant to verify the accuracy of the prediction for the flame front local velocity, that is, the relative velocity of the flow at point P (seen from the unburnt gases, that is, $u_{rel,P} = |\mathbf{u}_P - \mathbf{u}_{in}|$) defined by the intersection of the stoichiometric mixture fraction ($Z=Z_{st}$) level curve with the one for a given value of the scaled progress variable ($c=0.8$). Although this definition can lead to some oscillations due to transient effects, when comparing the flamelet model and the detailed chemistry it is also relevant to ensure that the FGM model is accurate in predicting unsteady effects. Temporal evolution for both $s_C$ and $u_{rel,P}$ are included in Fig.~\ref{fig:2D-GradZ_Speed}.

It is observed that the consumption speed increases when augmenting the mixing length, since the stretch effects, that include both curvature and strain along the flame front, are less intense and, therefore, the laminar flame speed of the unstreched flame is recovered. Additionally, note that the denominator of Eq.  (\ref{eq:fuel_consumption}) is computed along the inlet and does not account for the real arc length of the flame front which tends to increase with regards to $L_y$ when increasing the mixing length (see Fig. \ref{fig:2D-GradZ_ColormapMixingLengthTransition}). This is translated into an artificial increase of $s_C$ for high mixing lengths. The errors in $s_C$ for FGM are, after the transient evolution, reduced to less than 5$\%$.

With regards to the relative velocity at point P there exists a decrease with increasing mixing lengths due to the more accused divergence/convergence of the streamlines with higher $\delta_m$ as observed in the top right plot from Fig. \ref{fig:2D-GradZ_Speed}. Comparing both models, as the initial condition corresponds to a planar front, the transient evolution shows more oscillations for the higher mixing lengths since the streamlines that accommodate the flow \cite{Chung2007}, undertake a longer evolution. However, the differences between both models reduces to less than 5 $\%$ as the flame evolves.

This analysis demonstrates that the FGM is able to recover the flame structure and propagation of the flame when including preferential diffusion effects for a stratified flame with a limited variation in composition.

\subsection{Triple flame}\label{sec:2D-TripleFlame}

The final test case to examine the proposed preferential diffusion model is a triple flame configuration. A triple flame is the resulting flame structure generated by the propagation of a flame front in a partially premixed system, which originates from the mixture fraction gradient at the inlet. This flame structure is characterized by the formation of a triple point with a leading edge flame propagating with lean and rich premixed branches on the sides~\cite{Chung2007,VanOijen2004NumericalStudyConfined,Ray2000,Im1999}. Moreover, these flames interact with intermediate products to form a trailing diffusion flame. This complex flame structure is altered by the variation in local burning rates given by the diffusive fluxes of hydrogen on the flame front and brings an excellent validation case for the model. 

In this context, previous works have demonstrated the validity of FGM methods to predict the triple flame structure for specific values of mixing lengths using a unity Lewis approach \cite{VanOijen2004NumericalStudyConfined, Knudsen2012CapabilitiesLimitationsMultiregime}. It has been shown that when the mixture fraction gradient is low, the flame structure remains like a premixed flame, but as the gradient increases, the flame structure changes and departs from a premixed front \cite{ Knudsen2012CapabilitiesLimitationsMultiregime}. In this work, we extend this study to triple flames with preferential diffusion effects to determine the validity of the FGM under such conditions.

\begin{figure}
    \centering
    \includegraphics{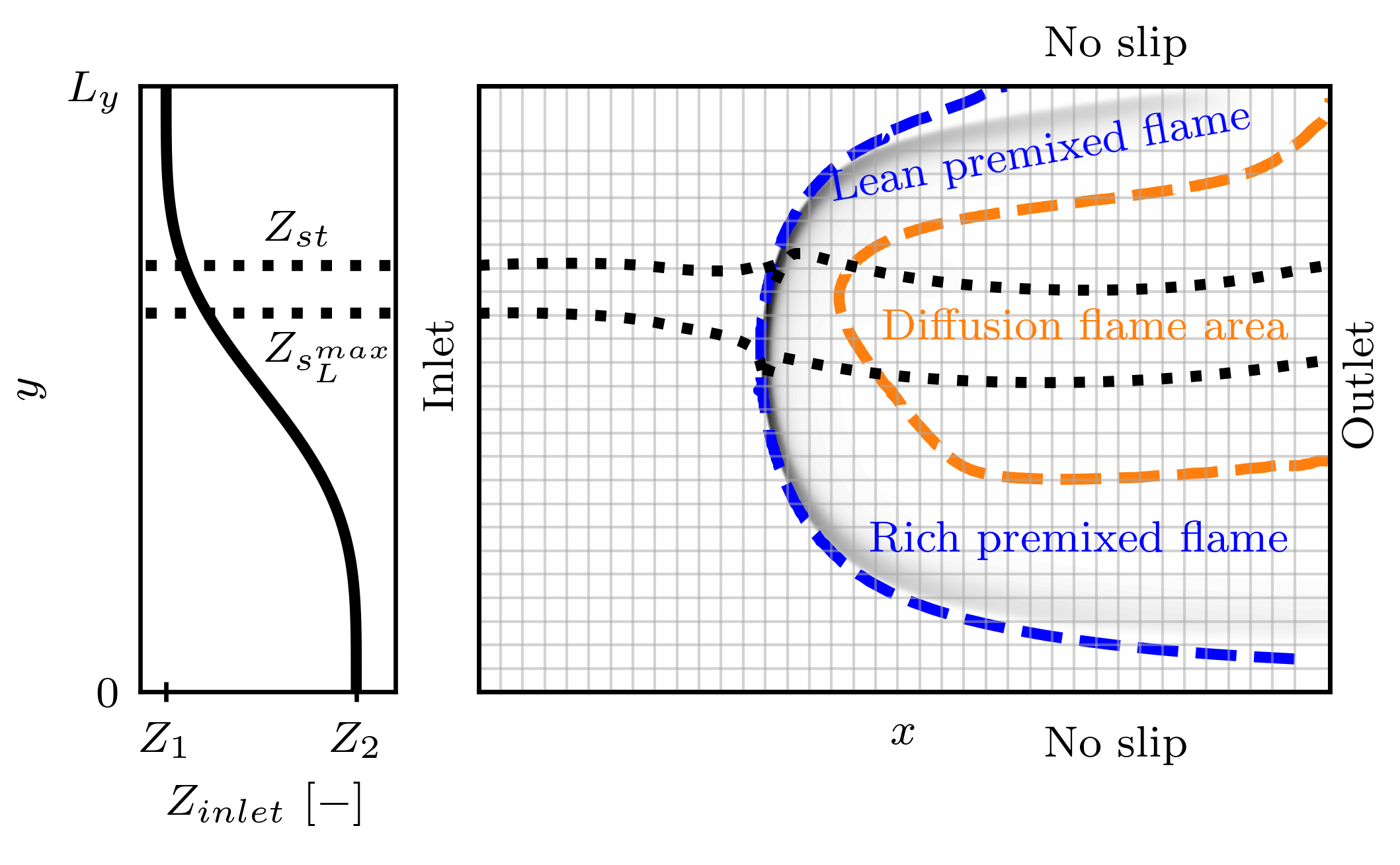}
    \caption{Triple flame schematics, indicating the injection profile of mixture fraction $Z_{in}(y)$ from Eq.\eqref{eq:2D-Triple_inlet_profile}. Flame structure zones determined from $I_{Takeno}$. Dashed black lines mark the position and level curves for stoichiometric and maximum one-dimensional laminar flame speed mixtures.
    }
    \label{fig:2D-Triple_schematics}
\end{figure}

The flame configuration in this hydrogen triple flame is shown in Fig. ~\ref{fig:2D-Triple_schematics}, which illustrates the flame shape for a mixture fraction inlet profile $Z_{in}(y)$ given by Eq. (\ref{eq:2D-Triple_inlet_profile}).

\begin{equation}
\label{eq:2D-Triple_inlet_profile}
Z_{\text{in}}=\frac{Z_{1}+Z_{2}}{2}+\frac{Z_{1}-Z_{2}}{2} \cos \left\{\frac{\pi}{2}\left[1-\cos \left(\frac{\pi y}{L_y}\right)\right]\right\} \quad \quad y \in \left[0,L_y \right]\text {. }
\end{equation}

Note that this inlet function is identical to the one used in the previous subsection~\ref{sec:2D-GradZ} but only taking half of the profile. The limits $Z_1$ and $Z_2$ have been chosen such that these mixture fractions cover most of the flammability range and have approximately the same laminar flame speed $s_L$ ($\approx 0.55$ m/s).
Moreover, different rectangular domains of dimensions $L_x$ and $L_y$ are defined (see Table \ref{tab:2D-Triple_table}) to consider different degrees of stratification. For all the cases, the minimum and maximum values of $Z_{in}$ are given by $Z_1=0.01445$ and $Z_2=0.1602$ that correspond to equivalence ratios of 0.5 and 6.5, respectively. From this profile, species mass fractions and enthalpy are obtained for the inlet conditions of the DC model. In this configuration, slip conditions are imposed at the top and bottom sides while Neumann conditions are used for the outlet boundary.

Finally, the same formula as in previous section, see Eq. (\ref{eq:2D-GradZ_inlet-mixing-length}), is used to define the mixing length::

\begin{equation}
\label{eq:2D-Triple_inlet_mix_length}
\delta_{\mathrm{m}}=(Z_2-Z_1)\left(\left.\frac{d Z_{in}}{d y}\right|_{y=L_y / 2}\right)^{-1} \text {. }
\end{equation}

The computational cases under investigation are presented in Table~\ref{tab:2D-Triple_table}. To initialize the simulation, fields for DC simulations are defined from a simple step-like function from $c=0$ to $c=1$ at the fixed distance $2/3 L_{x}$ for flames I and II and distance $4/5 L_x$ for flame III and the field $Z(x,y)|_{t=0}=Z_{in}(y)$ for mixture fraction in order to ensure the correct evolution of the flame without requiring long simulation times (such fields are used to define the species and enthalpy fields). 
Simulation is extended for $t_{1}=2\tau(\phi=1)$, where $\tau(\phi=1)$ is the characteristic time scale (see section \ref{sec:1D}) evaluated for the stoichiometric mixture, for a preliminary simulation with the DC model.
Then the resulting fields are used as the new initial condition for both simulations.

\begin{table}
    \centering
    \captionof{table}{Triple flame simulation parameters.}\label{tab:2D-Triple_table}
    \begin{tabular}{|c|c|c|c|}
    \hline
    Case & $\delta_m$/$l_F(\phi=1))$ & $L_y$ (mm)   &   $L_x$ (mm)\\ \hline
    I & 3.02	 & 2.5   &   7.5      \\ \hline
    II & 6.05  & 5   &  15   \\ \hline
    III &  9.7  & 8  &  25 \\ \hline
    \end{tabular}
\end{table}

\begin{figure}
    \centerline{\includegraphics[width=15.0cm]{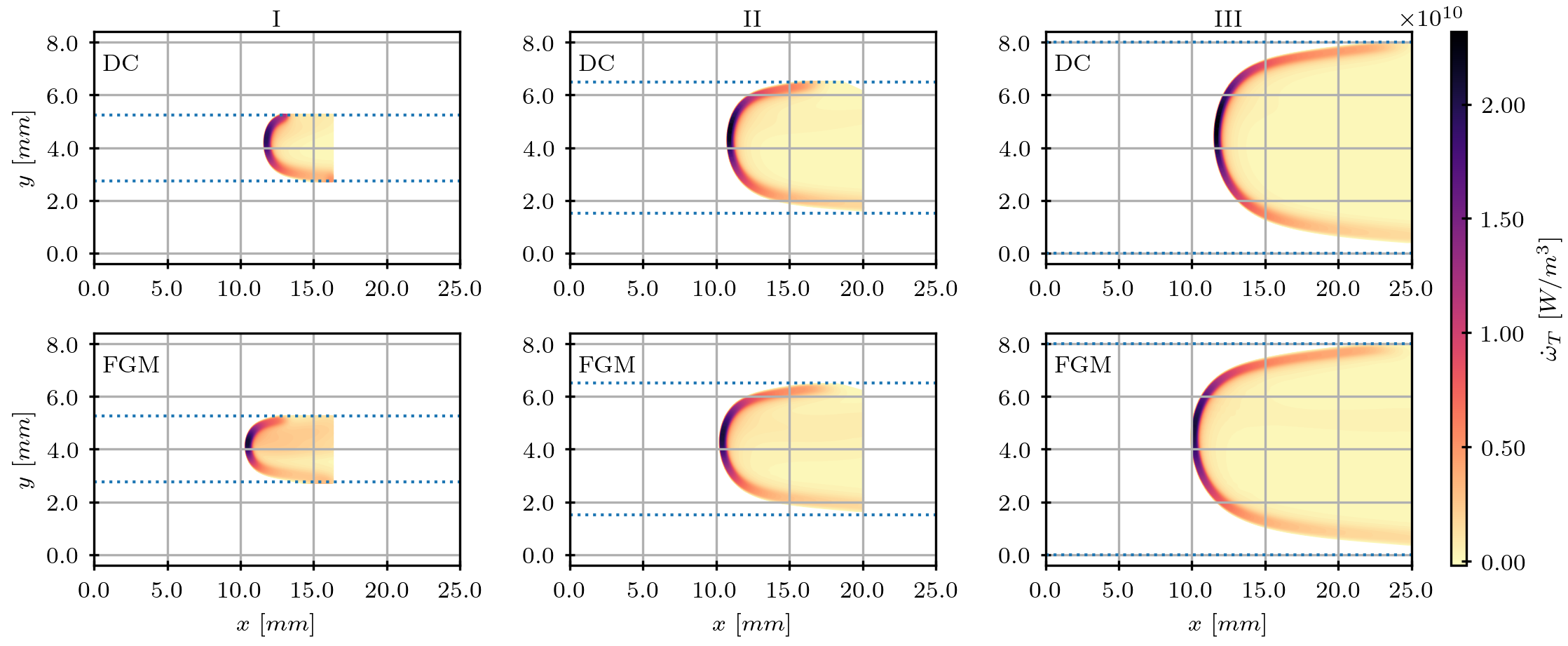}}
    \caption{Contour plots of $\dot{\omega}_{T}$ for the triple flames defined by Table \ref{tab:2D-Triple_table}.
    From left to right, the times of the snapshots are $t=1.58$, $1.58$ and $3.17$ ms. Each column presents the corresponding mixing length from the table, with the results from the DC model at the top panel and the FGM model at the bottom panel. All figures are centered for visualization. Dotted blue lines delimit the physical domain of the simulations.}\label{fig:2D-TripleFlame_heat_release}
\end{figure}

The heat release field of the different cases is shown in Fig.~\ref{fig:2D-TripleFlame_heat_release} in order to understand the influence of the mixing length on the flame dynamics and thermochemical structure (see Table~\ref{tab:2D-Triple_table}). Some discrepancy is evident near the slip condition at the top and bottom sides where the DC model exhibits shorter tails on both the lean and rich sides of the flame. It is shown that this effect diminishes as the mixing length increases. Also a less curved flame is obtained for the DC model, fact that is evident for high mixture fraction gradients (flame I). The variation of the flame curvature with the mixing length is in agreement with results from literature involving triple flames under unity Lewis number assumption \cite{VanOijen2004NumericalStudyConfined, Illana2021ExtendedFlameIndex}.

\begin{figure*}[h]
    \centerline{\includegraphics{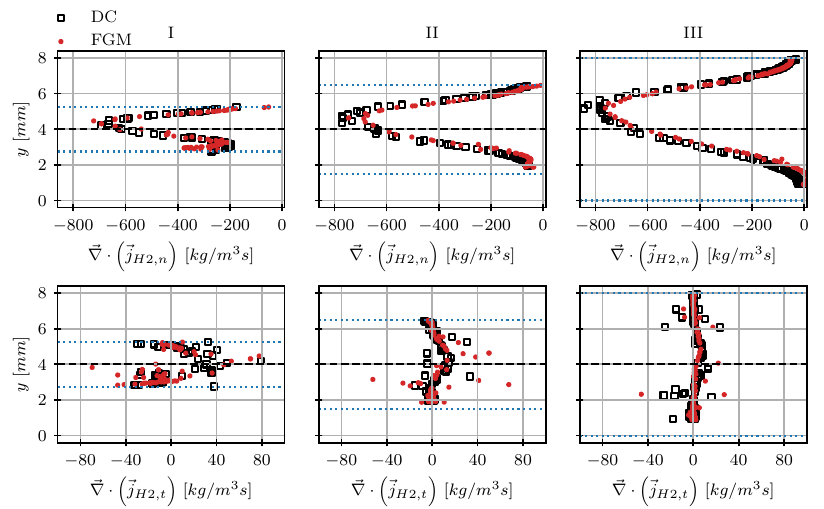}}
    \caption{Values for the normal (top subplots) and tangential (bottom subplots) diffusive flux balances for $\mathrm{H_2}$ for all the flames from Table \ref{tab:2D-Triple_table}. All plots are centered for visualization. Blue dotted lines delimit their physical domains. For each subplot, the values for the DC model are shown with black empty squares while the FGM model with red dots.}\label{fig:2D-TripleFlame_diffusive_flux}
\end{figure*}

Figure~\ref{fig:2D-TripleFlame_diffusive_flux} shows the balances for the normal and tangential diffusive fluxes for $\mathrm{H_2}$ as defined in section \ref{sec:2D-GradZ}. On the one hand, the diffusive flux balance in the normal direction is well predicted by the FGM method for each mixing length. On the other hand, the tangential diffusive flux balance contributions are magnified close to the centerline and the boundaries when the mixing length is reduced. The values are at least one order of magnitude smaller than those corresponding to the normal fluxes. In summary, even for the shortest mixing lengths, the FGM model provides an excellent agreement with the DC model.

\begin{figure*}[h]
    \centerline{\includegraphics{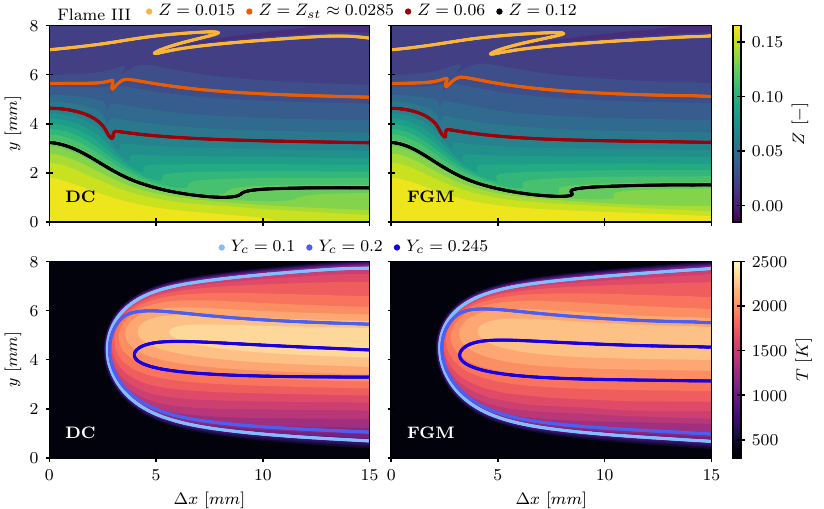}}
    \caption{Contour plot of $Z$ (top) and $T$ (bottom) for the simulation of triple flame III. Results from the $DC$ model are represented in the left column while the $FGM$ model results at the right one.  Additionally, figures for $Z$ (top) include the iso-lines for values $Z=0.015$ (light orange), $Z_{st}=0.0285$ (orange), $0.06$ (brick red) and $0.12$ (black). Figures for temperature include iso-lines for $Y_c$ with values $Y_c=0.1$ (light blue), $0.2$ (blue) and $0.245$ (dark blue).}\label{fig:2D-TripleFlame_ColormapMultiple}
\end{figure*}

A visualization of the flame in this configuration can be distinguished in Fig.~\ref{fig:2D-TripleFlame_ColormapMultiple} for flame III (the lowest $Z$ gradients) for both models. The plot includes comparisons for control variable $Z$ and temperature. Iso-lines for mixture fraction and progress variable have also been included for reference. The plots show a remarkable good agreement of the flamelet model with the reference DC solutions with minor differences only detected close to the top and bottom boundaries.

Moreover, the model predicts the edge flame located at the leading edge of the triple flame, with also relevant predictions of the rest of quantities. The flamelet manifold can capture, then, the complex structure appearing in a triple including the variations in burning rates. Indeed, it can be observed that the FGM predicts the position of the stabilization point (maximum velocity point), which corresponds to a rich mixture fraction of $Z \approx 0.046$ ($\phi\approx 1.65$) in all the cases \cite{Im1999}. Recall that the maximum flame speed for a one-dimensional hydrogen-air flame is found at $\phi = 1.65$. A small but noticeable difference in temperature is captured in the burnt mixture region for $Z_{st}$, where the DC model shows slightly higher temperatures along the 
trailing diffusion flame. Such differences are due to the fact that the manifold is constructed from one-dimensional premixed flames and, therefore, some differences are expected in the diffusion flame region.

\begin{figure*}[h]
    \centerline{\includegraphics{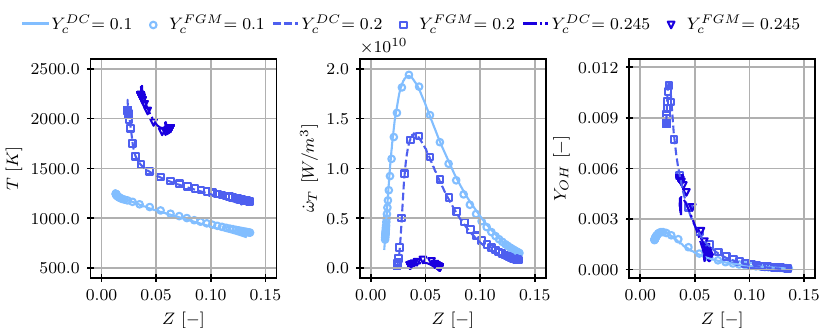}}
    \caption{Values of $T$, $\dot{\omega}_T$ and $Y_{\mathrm{OH}}$ vs $Z$ subject to isolines of values of $Y_{c}=0.1$ (light blue), $0.2$ (blue) and $0.245$ (dark blue) for DC model (lines) and FGM (empty markers) for flame III. These isolines are shown in the temperature map from figure \ref{fig:2D-TripleFlame_ColormapMultiple}.}\label{fig:2D-Triple_isoYc}
\end{figure*}

\begin{figure*}[h]
    \centerline{\includegraphics{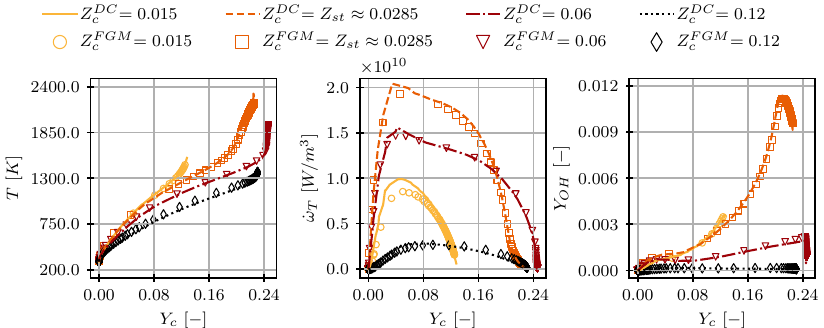}}
    \caption{Values of $T$, $\dot{\omega}_T$ and $Y_{\mathrm{OH}}$ vs $Y_{c}$ subject to isolines of values of $Z = 0.015$ (light orange), $Z_{st}=0.028512$ (orange), $0.06$ (dark red) and $0.12$ (black) for DC model (lines) and FGM (empty markers) for flame III. These isolines are shown in the mixture fraction map from figure \ref{fig:2D-TripleFlame_ColormapMultiple}.}\label{fig:2D-Triple_isoZ}
\end{figure*}

A more quantitative analysis of the flame structure for flame III is shown in Figs. \ref{fig:2D-Triple_isoYc} and \ref{fig:2D-Triple_isoZ}. Profiles for $T$, $\dot{\omega}_{T}$ and $Y_{\mathrm{OH}}$ are represented along the level curves of progress variable $Y_c$ and mixture fraction $Z$ included in Fig. \ref{fig:2D-TripleFlame_ColormapMultiple}. The results show that there exists a remarkable agreement between the results provided by the DC and FGM models with only slight differences in the very lean mixtures for the heat release. Such differences can be produced by the influence of the boundary conditions. This comparison demonstrates that the FGM accurately predicts the internal structure of the triple flame. As observed previously, a small difference in temperature predicted by the FGM model is also detected at the stoichiometric mixture fraction (Fig. \ref{fig:2D-Triple_isoZ}) close to the equilibrium due to the trailing diffusion flame.

To support the analysis and facilitate the identification of flame regimes, the Takeno index \cite{Yamashita1996NumericalStudyFlame} is used here to separate the combustion regimes ~\cite{Domingo2005DNSAnalysisPartially, Patel2008SimulationSprayTurbulence}. The index is computed from the scalar product of gradients of the fuel ($\nabla Y_{F}$) and oxidiser ($\nabla Y_{O}$) \cite{Yamashita1996NumericalStudyFlame} and is used to discriminate the premixed and diffusion flame zones of the DC model. It is given by:

\begin{equation}\label{eq: Takeno} 
    I_{FO} = \frac{\nabla Y_{F} \cdot \nabla Y_{O}}{|\nabla Y_{F} \cdot \nabla Y_{O}|}, \quad \quad \nabla Y_{F} \cdot \nabla Y_{O} \neq 0.
\end{equation}

\begin{figure*}[h]
    \centerline{\includegraphics[width=10.0cm]{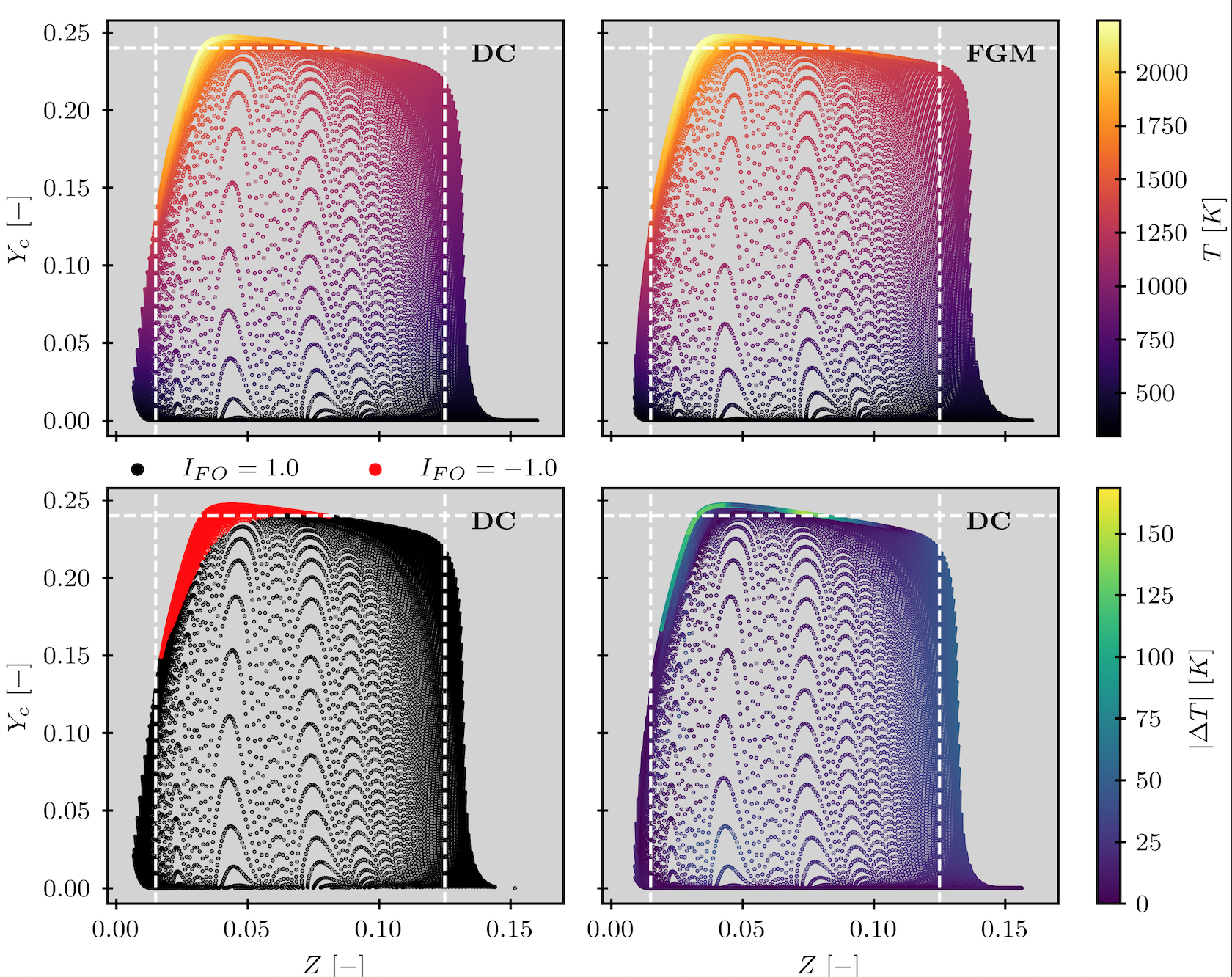}}
    \caption{Scatter plots of composition space $(Z,Y_{c})$ for the triple flame III from Sec.~\ref{sec:2D-TripleFlame}. Top figures: temperature coloured scatter plots for DC (left) and FGM (right). Bottom figures: scatter plots of DC colored by Takeno's index $I_{FO}$ (left) and temperature difference between models $|\Delta T|=|T^{DC} - T^{FGM}|$ in the areas where both spaces overlap (right). White dashed lines help the visualization to identify differences between composition space for DC and FGM models.}\label{fig:2D-TripleFlame_ErrorTakeno}
\end{figure*}

The index features premixed combustion by $I_{FO} > 0$ and diffusion flames by $I_{FO} < 0$.
Scatter plots of different quantities in composition space $(Z, Y_c)$ for flame III are shown in Fig. \ref{fig:2D-TripleFlame_ErrorTakeno}. The top subplots show the values of temperature for DC and FGM models, allowing to identify model discrepancies. As delimited by the rightmost white dashed lines, the FGM model predicts higher values of $Z$ and $Y_c$ than its DC model counterpart (region of $Z>0.12$). Inside the common $(Z, Y_{c})$ region between models, the value of $T^{DC} - T^{FGM}$ is shown in the bottom right subplot of the figure. Two areas are identified with magnified errors. On the one hand, the rightmost area increases its error as the model gets outside the shared composition space. On the other hand, there exists a second area between $0.02\leq Z \leq 0.09$ and maximum values of $Y_{c}$, identified in the upper left area of the scatter plot, where errors are detected. This area corresponds to the surroundings of the values of burnt $Z=Z_{st}$, where the trailing diffusion flame is observed, as seen in the schematic of Fig.~\ref{fig:2D-Triple_schematics}, and identified by the negative values of $I_{FO}$ in the bottom left plot. Therefore, this is the region where the Takeno index transitions from premixed to diffusion-dominated combustion and the flamelet model, constructed from premixed flames, shows some limitations in the prediction of the flame structure.

\begin{figure*}[h]
    \centerline{\includegraphics{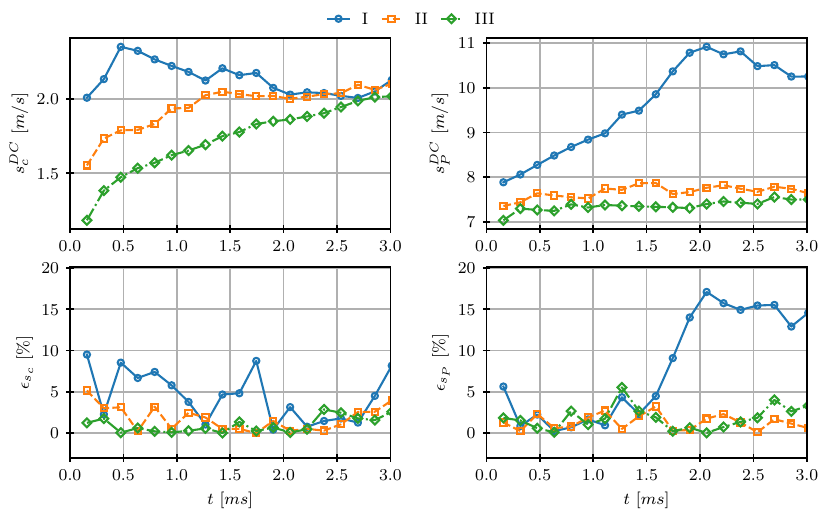}}
    \caption{Speed comparison for the simulations of the triple flame. Top left: consumption speed $s_{C}$ for DC model. Top right: relative velocity $u_{rel,P}$ for point P for DC model. Bottom left: relative difference $\epsilon_{s_C} = |(s_C^{FGM} - s_C^{DC}) / s_C^{DC}|$ between models. Bottom right: relative difference $\epsilon_{u_{rel,P}} = | (s_P^{FGM} - s_P^{DC}) / s_P^{DC} |$ between models.}\label{fig:2D-TripleFlame_Speed}
\end{figure*}

The final comparison between the flamelet model and detailed chemistry is made in Fig.~\ref{fig:2D-TripleFlame_Speed} in terms of consumption speed and the relative velocity at the leading edge (see definitions in section \ref{sec:2D-GradZ}). This comparison is made in order to ensure that the global burning rates are well recovered by the FGM method. The agreement for the consumption speed $s_C$ between models confirms that the flame structure is well-reproduced by the FGM model (relative errors smaller than 5$\%$). The relative velocity for the leading edge point P is, in general, well-captured even some differences are observed for the smallest mixing length (flame I). Such differences are attributed to slight discrepancies in flame structure in regions of high gradients. Notwithstanding, in general, the analysis of the leading edge suggests that the FGM model predicts the local burning rates and the correlation with the DC solutions is satisfactory.

\section{Conclusions and future works}

In this work, a method has been derived to account for differential diffusion effects through the mixture-averaged model in the FGM combustion model based on \cite{deSwart2010,Donini2015}. The contributions from the molecular weight and the velocity correction to assure mass conservation have been retained in the formulation. Moreover, a full formulation of the model has been implemented which retains all the cross terms. A complete description has been given for the theoretical framework, focusing on the numerical aspects and the interpolation algorithm to obtain the coefficients for the manifold with minimum noise at a relatively low computational cost.

Several canonical configurations have been considered and compared to the reference simulations to quantify the method robustness and consistency. First, one-dimensional flames have been reproduced with the FGM  and compared with Cantera's solver showing that the FGM model is capable of predicting the flame speed, thickness and internal structure with minimal deviations. In the second step, an adiabatic bidimensional stratified flame with a limited variation in composition has been solved with the FGM and compared with the finite rate. Flames with several mixing lengths have been simulated to evaluate the influence of the gradients and curvature on the flame structure and propagation. From the results, it was observed that the flame was essentially well-reproduced and only some differences were observed in specific regions close to the boundaries. Also the consumption speed and the velocity of propagation were properly predicted. Finally, the ability of the model has been assessed in a triple flame configuration where the mixture fraction encompass the entire flammability range. Similarly to previous case, several mixing lengths have been tested. It has been shown that, even for this complex structure, the FGM model with mixture-averaged is capable of predicting the flame structure and the flame propagation. It has been shown that for all the configurations the contributions from the tangential fluxes, even enhanced when the decreasing the mixing length, are relatively small in comparison to the normal one. Therefore, the deviations from the internal flame structure contained in the low-dimensional manifold are limited.

It is concluded, then, that the small overcost to obtain and save the coefficients $\Gamma_{\mathcal{Y}_i,\mathcal{Y}_j}$ together with the capabilities of the FGM with mixture-averaged transport to reproduce these canonical flame demonstrate its suitability and potential for the prediction and application to general flame configurations.

In future works, the method will be extended to incorporate heat losses. Also the extension to turbulent combustion regime to consistently account for fluctuations and apply the method to the simulation of turbulent hydrogen flames is a natural step to be worked out.

\section{Acknowledgements}

The research leading to these results has received funding from the European Union's Horizon 2020 Programme under the CoEC project, grant agreement No. 952181, and HyInHeat project grant agreement No. 101091456, AHEAD PID2020-118387RB-C33 and ORION TED2021-131618B-C22 projects from the Ministerio de Ciencia e Innovaci\'on. EMFS acknowledges the grant Ajuts Joan Oró per a la contractació de personal investigador predoctoral en formació (FI 2023) cofinanced by the EU, Generalitat de Catalunya: Departament de Recerca i Universitats and Agencia de Gestión de Ayudas Universitarias y de Investigación (AGAUR). DM acknowledges the grant Ayudas para contratos Ramón y Cajal (RYC 2021): RYC2021-034654-I from Ministerio de Ciencia e Innovación. 

\appendix
\section{1D Flames additional results}\label{appendix:1D}

This appendix details the results and analysis obtained for the validation of the quasi-one-dimensional flames. 

\subsection{Laminar flame speed and thermal flame thickness}\label{appendix:1D_sub1}

\begin{figure}[h]
    \centering
    \centerline{\includegraphics{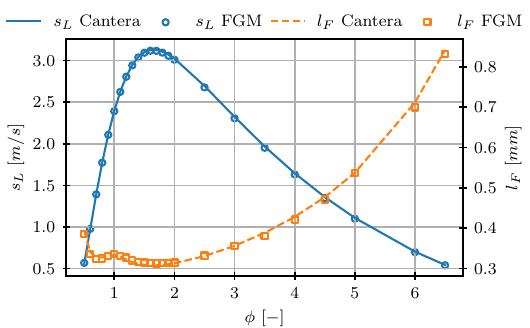}}
    \caption{Flame speed $s_L$ and thermal flame thickness $l_F$ comparison between Cantera (solid and dashed lines respectively) and Alya FGM (circle and square markers respectively) with control variables $(\mathcal{Y}_{1},\mathcal{Y}_{2})=(Y_{\mathrm{H_2O}},Z)$ for one-dimensional adiabatic hydrogen/air flames at atmospheric pressure and unburnt gas temperature of 298.15 K.}
    \label{fig:flame_speed}
\end{figure}

The comparisons of laminar flame speed $s_L$ and thermal flame thickness $l_F$ between Cantera and the proposed model in the code Alya are shown in Fig. ~\ref{fig:flame_speed}.

Table~.\ref{table:1D-results} shows the values of the laminar flame speed and thermal flame thickness from Eqs. \eqref{eq:laminar_flame_speed} and \eqref{eq:thermal_flame_thickness} for Cantera (reference value) and Alya with FGM and mixture-averaged diffusion model. For the flame speed, the values are in perfect agreement with a relative error $\epsilon_{s_L} = | (s_L^{FGM} - s_L^{C}) / s_L^{C} |$ that oscillates between 0.1 and 1.56$\%$ while for the flame thickness the relative error given by $\epsilon_{l_F} = | (l_F^{FGM} - l_F^{C}) / l_F^{C} |$ ranges between 0.13 and 2.35$\%$. As it can be seen from the flame speed results, there is a slight reduction in accuracy when enriching the mixture, fact that is consistent with the deterioration of $Y_{\mathrm{H_2O}}$ as progress variable for fuel rich mixtures. Regardless, even at $\phi=6.5$ the error remains very low, proving the accuracy of the method.

\begin{table}[h]
\centering
\captionof{table}{Flame speed and laminar flame thickness values comparison.} \label{table:1D-results}
\begin{tabular}{lrrrrrr}
\toprule
 & $\underset{[m/s]}{s_L^{FGM}}$ & $\underset{[m/s]}{s_L^{CT}}$ & $\underset{[mm]}{l_F^{FGM}}$ & $\underset{[mm]}{l_F^{CT}}$ & $\epsilon_{s_L}$ & $\epsilon_{l_F}$ \\
$\phi$ $[-]$ &  &  &  &  &  &  \\
\midrule
0.50 & 0.5691 & 0.5603 & 0.3849 & 0.3942 & 1.5652 & 2.3573 \\
0.60 & 0.9778 & 0.9733 & 0.3362 & 0.3403 & 0.4586 & 1.1910 \\
0.70 & 1.3922 & 1.3934 & 0.3238 & 0.3272 & 0.0846 & 1.0401 \\
0.80 & 1.7731 & 1.7770 & 0.3243 & 0.3273 & 0.2198 & 0.9321 \\
0.90 & 2.1073 & 2.1110 & 0.3311 & 0.3315 & 0.1755 & 0.1346 \\
1.00 & 2.3915 & 2.3940 & 0.3353 & 0.3341 & 0.1040 & 0.3361 \\
1.10 & 2.6237 & 2.6264 & 0.3309 & 0.3296 & 0.1025 & 0.3998 \\
1.20 & 2.8062 & 2.8100 & 0.3266 & 0.3219 & 0.1334 & 1.4394 \\
1.30 & 2.9423 & 2.9476 & 0.3196 & 0.3157 & 0.1801 & 1.2357 \\
1.40 & 3.0381 & 3.0429 & 0.3158 & 0.3113 & 0.1582 & 1.4231 \\
1.50 & 3.0946 & 3.1007 & 0.3142 & 0.3087 & 0.1952 & 1.7697 \\
1.60 & 3.1183 & 3.1265 & 0.3138 & 0.3075 & 0.2638 & 2.0474 \\
1.70 & 3.1172 & 3.1268 & 0.3120 & 0.3075 & 0.3056 & 1.4451 \\
1.80 & 3.0956 & 3.1068 & 0.3127 & 0.3085 & 0.3592 & 1.3844 \\
1.90 & 3.0578 & 3.0715 & 0.3130 & 0.3102 & 0.4438 & 0.9144 \\
2.00 & 3.0103 & 3.0241 & 0.3142 & 0.3125 & 0.4563 & 0.5631 \\
2.50 & 2.6775 & 2.6949 & 0.3316 & 0.3304 & 0.6460 & 0.3710 \\
3.00 & 2.3058 & 2.3237 & 0.3549 & 0.3559 & 0.7687 & 0.2721 \\
3.50 & 1.9523 & 1.9674 & 0.3810 & 0.3881 & 0.7688 & 1.8281 \\
4.00 & 1.6313 & 1.6464 & 0.4208 & 0.4278 & 0.9151 & 1.6226 \\
4.50 & 1.3486 & 1.3599 & 0.4706 & 0.4766 & 0.8252 & 1.2595 \\
5.00 & 1.1003 & 1.1095 & 0.5373 & 0.5371 & 0.8337 & 0.0323 \\
6.00 & 0.6978 & 0.7038 & 0.6995 & 0.7116 & 0.8490 & 1.7027 \\
6.50 & 0.5397 & 0.5440 & 0.8314 & 0.8418 & 0.7870 & 1.2378 \\
\bottomrule
\end{tabular}
\end{table}

\subsection{Study on the $\Gamma^{'}_{i,j}$ coefficients}\label{appendix:1D_coeffs}

\begin{figure*}[h]
    \centerline{\includegraphics{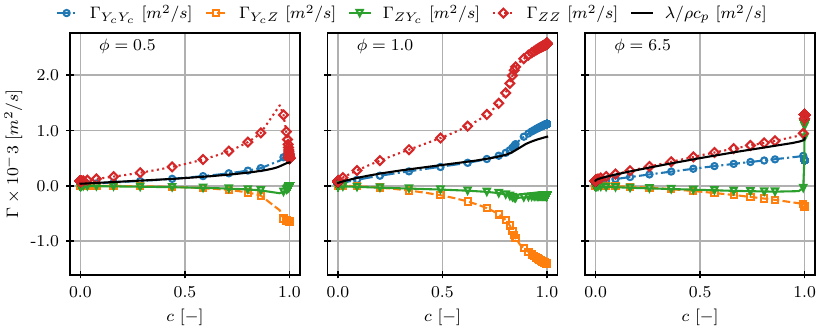}}
    \caption{Evolution of the coefficients $\Gamma_{Y_{c}, Y_{c}}$ (blue), $\Gamma_{Y_{c}, Z}$ (orange), $\Gamma_{Z, Y_{c}}$ (green) and $\Gamma_{Z, Z}$ (red) as a function of the normalized progress variable along the flames for the equivalence ratios of $0.5$ (left), $1.0$ (center) and $6.5$ (right). Results for one-dimensional adiabatic hydrogen/air flames at atmospheric pressure and unburnt gas temperature of 298.15 K.
    }\label{fig:fgm_mixing_coefficients}
\end{figure*}

Finally, the analysis of the 1D flames is concluded by studying the tabulated transport coefficients in the phase
space. For this purpose, Fig.~\ref{fig:fgm_mixing_coefficients} shows the evolution of the coefficients $\Gamma_{Y_c,Y_c}$, $\Gamma^{'}_{Y_c,Z}$, $\Gamma^{'}_{Z,Y_c}$ and $\Gamma_{Z,Z}$ appearing in Eqs. (\ref{eq_solve_Yc}) and (\ref{eq_solve_Z}) as a function of the the reactive space $c$ along the one-dimensional flames. The values for $\Gamma_{Y_{c},Y_{c}}$ and $\Gamma_{Z,Z}$ remain positive with similar values to the thermal diffusivity $D_{th}$, and therefore, they act as diffusion coefficients for $Y_c$ and $Z$, respectively. The remaining terms $\Gamma^{'}_{Y_{c},Z}$ and $\Gamma^{'}_{Z,Y_{c}}$ are of negative sign, and, therefore, the contributions $\nabla \cdot (\rho \Gamma^{'}_{Y_c,Z} \, \nabla Z)$ and $\nabla \cdot (\rho \Gamma^{'}_{Z,Y_c} \, \nabla Y_c)$ cannot be directly associated to diffusive fluxes without further analysis. In fact, the term $\nabla \cdot (\rho \, \Gamma^{'}_{Z,Y_c} \, \nabla Y_c)$ acts as a source term since it causes the variation of mixture fraction across the flame. Because of the choice of water as progress variable and, the fact that its Lewis number
is very close to unity, the values of $\Gamma_{Y_{c},Y_{c}}$ and $D_{th}$ are almost identical.

Near equilibrium ($c \approx 1$), the coefficients show a sharp gradient that increases with $\phi$. This can be explained by the selection of the progress variable ($Y_c= Y_{\mathrm{H_2O}}$), which evidences the limitations of this progress variable definition for fuel-rich conditions. The identification of an optimal progress variable is outside the scope of this work, and the reader is referred to more specific studies on this topic for a description of optimization strategies \cite{Niu2013OptimizationbasedApproachDetailed, Efimov2018FGMREDxChemically}. Hence, a simple definition of progress variable is retained here, since it reproduces well most of the conditions, has a simple description and facilitates the analysis.

\clearpage

\bibspacing=\dimen 100
\bibliographystyle{plain}
\bibliography{Perez_Sanchez_FGM_preferential_diffusion_hydrogen}

\end{document}